\documentclass[authoryear,12pt]{elsarticle}
 \usepackage{graphicx}

\usepackage{amssymb}
\usepackage{amsmath}
\usepackage{ctable} 
\usepackage{soul}
\usepackage{subfig}

\makeatletter
\def\ps@pprintTitle{%
  \let\@oddhead\@empty
  \let\@evenhead\@empty
  \let\@oddfoot\@empty
  \let\@evenfoot\@oddfoot
}
\makeatother

\begin{document}

\begin{frontmatter}

\title{Joint fitting reveals hidden interactions in tumor growth}

\author[ifly,dieudone]{L. Barberis}
\address[ifly]{Instituto de F\'isica de L\'iquidos y Sistemas Biol\'ogicos \\ CONICET - UNLP, La Plata, Argentina. }
\ead{lbarberi@iflysib.unlp.edu.ar}

\author[inifta]{M. A. Pasquale}
\address[inifta]{Instituto de Investigaciones Fisicoqu\'imicas Te\'oricas y Aplicadas \\ CONICET - UNLP, La Plata, Argentina. }

\author[ifeg,famaf]{C. A. Condat}
\address[ifeg]{Instituto de F\'isica Enrique Gaviola - CONICET - UNC, C\'ordoba, Argentina. }
\address[famaf]{Facultad de Matem\'atica Astronom\'ia y F\'isica, UNC, C\'ordoba, Argentina.}

\address[dieudone]{Laboratoire J. A. Dieudonn\'e, Universit\'e de Nice Sophia Antipolis, Nice , France.}

\begin{abstract}
Tumor growth is often the result of the simultaneous development of two or more cancer cell populations. Their interaction between them characterizes the system evolution. To obtain information about these interactions we apply the recently developed vector universality (VUN) formalism to various instances of competition between tumor populations. The formalism allows us: (a) to quantify the growth mechanisms of a HeLa cell colony, describing the phenotype switching responsible for its fast expansion, (b) to reliably reconstruct the evolution of the necrotic and viable fractions in both \emph{in vitro} and \emph{in vivo} tumors using data for the time dependences of the total masses, and (c) to show how the shedding of cells leading to subspheroid formation is beneficial to both the spheroid and subspheroid populations, suggesting that shedding is a strong positive influence on cancer dissemination.

\end{abstract}
 
\end{frontmatter}


\section{Introduction}\label{s_intro}

The development of multiple mathematical models in tumor biology results from their potential to describe and predict neoplastic growth and the effect of applied therapies. These models use both phenomenological and mechanistic descriptions of growth, whose success requires an understanding of cellular self-organization during the progress of the disease. In particular, empirical models are based on the observation that tumor growth results from cellular multiplication concomitant with processes that, in most cases, limit the size of the system.

Growth is described by extensive variables such as volume, mass, length, or number of cells, represented  as functions of time. The growth of individual organisms \citep{Laird1965}, tumors \citep{Bajzer1996}, and other biological systems \citep{Zwietering1990} is often well fitted by sigmoidal functions. Widely used empirical descriptions are those of Gompertz  \citep{Fujikawa1989} and von Bertalanffy-West \citep{West2001}; under certain conditions, Malthusian, i.e. exponential,  functions are also used \citep{Zwietering1996}. These growth curves are very useful to interpret and measure various physical properties of the tumor, such as necrosis  diameter and the volume increase rate in experimental systems \citep{Mueller2000}, the pre-angiogenic network structure \citep{Guiot2006}, and the cell-cycle fraction as a function of time \citep{Jiang2005}, among others. Furthermore, there is a manifold of ad hoc models \citep{Marusic1994,Mombach2002,deVladar2006,Castorina2006,Gonzalez2006,Delsanto2008a} that try to 
justify and 
unify these empirical 
descriptions in an attempt to reveal their common features.

Identifying correlations among phenomenological growth curves can shed light on the underlying biological processes responsible for tumor growth. It is therefore important to get accurate fitting functions in order to recover the greatest possible amount of information about the system. Since obtaining reliable data on growth is a difficult, onerous, and tedious task, the ability to recover information from a limited amount of data is an important goal for modelers.  It is with this objective that we have developed a powerful fitting tool \citep{Barberis2011}, which can provide accurate fitting functions for diverse systems consisting of two or more individuals or populations growing together. As a bonus, the method is able to evince hidden interactions among the populations without any previous assumptions on their nature.

Except in its earliest stages, cancer is usually a multipopulation system: mutations appear and sometimes prevail, cells die and generate necrotic volumes, and even the cells in a primary tumor may compete with those in the colonized tissue. In this paper we show how information about cancer cell {\bf interactions} can be extracted from growth curves. To do this we will consider diverse experimental systems: jointly growing cancer cell phenotypes, \emph{in vitro} multicellular tumor spheroid growth (where live and dead cells may be thought of as competing), \emph{in vivo} experimental implanted tumors, and a system in which spheroids and subspheroids evolve simultaneously.

Although fitting functions to cancer growth curves is a time-honored procedure to glean information about tumor properties, the VUN method goes far beyond, enabling us to extract information from the correlations between growth curves. This is done for the first time in this paper, where we investigate the correlated growth of two cancer cell subpopulations to characterize how their mutual interactions and their interactions with the environment affect their growth.

In Section \ref{s_VUN} we briefly describe the Phenomenological Universalities (PUN) formalism introduced by \cite{Castorina2006} and extended to competing organisms and populations by \cite{Barberis2011} and \cite{Barberis2012}. Examples of the application of the PUN analysis  to the aforementioned experiments are  provided in Section \ref{s_results}, where we apply the procedure to several cancer growth data sets. A discussion of the results is presented in Section \ref{s_discussion}, where we comment on the applicability and limitations of the method.

\section{Vector Universalities}\label{s_VUN}

The universal growth formalism for one-population systems is a systematic generalization, due to Delsanto's group \citep{Castorina2006}, of the well-known growth functions of Malthus, Gompertz, and von Bertalanffy. In the following a brief description of the method is presented, starting with the one-population formalism (scalar equations) and then proceeding to its vector extension, which is suitable to describe the joint growth of two or more populations.

\subsection{One-population growth equations.}

A clear and comprehensive characterization of the growth functions  for one-population systems was presented by \cite{deVladar2006}, who described growth
using two first-order differential equations, one for the population size
$y(t)$,

          \begin{equation}\label{e_intro_scalargrowth}
           \dot{y}(t)=a(t)y(t),
          \end{equation}
(the \emph{growth} equation) and another for the growth rate $a(t)$,

        \begin{equation}\label{e_intro_scalarrate}
           \dot{a}(t)=\left[\theta a(t)-\rho\right]a(t),
          \end{equation}
(the \emph{rate} equation), where $\theta$ and $\rho$ are two real parameters.
Various combinations of these parameters reproduce, among others, the
$\theta$-logistic, von Bertalanffy, Gompertz, and potential growth equations. \cite{deVladar2006} indicates that the size of the dimensionless parameter $\theta$ defines
the density scale at which the reproduction rate of an individual is affected by
its interaction with the population, while $\rho^{-1}$ is a characteristic time
over which the individual downregulates its reproduction rate.

In the work of \cite{Castorina2006}, the right-hand side of
equation (\ref{e_intro_scalarrate}) is replaced by a power series expansion in the
rate $a(t)$,  

\begin{equation}\label{e_intro_scalarpower}
\dot{a}(t)=\sum_{m=0}^\infty \alpha_m a^m(t).
\end{equation}

Of course, by truncation and a suitable choice of the parameters $\alpha_m$,
Eq. (\ref{e_intro_scalarpower}) reduces to the different possibilities generated
by Eq. (\ref{e_intro_scalarrate}), but \cite{Castorina2006} suggest that
the concept of Phenomenological Universalities may be used in conjunction with
Eqs. (\ref{e_intro_scalargrowth}) and (\ref{e_intro_scalarpower}) as a tool for the
classification and interpretation of observed data in the context of
cross-disciplinary research. In fact, they have applied this concept to fields
as diverse as those of elastodynamics \citep{Pugno2008}, human growth
\citep{Delsanto2008a}, and cell proliferation in cancer \citep{Gliozzi2010}. 
In \cite{Barberis2010}, the use of complex variables $y(t)$ and $a(t)$ made
it possible to investigate the simultaneous variations of two phenotypic
features of an individual. This procedure showed the existence of correlations
between changes in the fat distribution of the human body. The behavior of oscillatory coupled systems was also studied with this method \citep{Delsanto2011}.

\subsection{Two or more populations: Vector formalism.}

We are now interested in the description of the correlations between variations in the same trait of different agents. As in the problem treated by \cite{Castorina2006}, the resulting generalization of the Phenomenological Universalities formulation is especially useful in those cases for which no reliable model is available.

We describe the time evolution of a given phenotypic feature observed in $n$
interacting agents through an $n$-component \emph{growth vector} $Y(t)$, and
postulate that the evolution of this vector is determined by a generalization of
the autonomous growth equation proposed by \cite{Castorina2006},

\begin{equation}\label{e_dotY}
\dot{Y}(t)=A\,Y(t),
\end{equation}
where $t$, the time, is a real continuous parameter, $Y(t) \in \mathbb{R}^n$,
and the dynamic operator $A[Y(t)]\in \mathbb{R}^{n\times n}$. According to the
PUN formulation, we assume that the rate of change of the functional $A[Y(t)]$
can be expressed as a power series expansion in the operator $A$ itself, which
we truncate to order $N$,

\begin{equation}\label{e_intro_dotA}
 \dot{A}(t)=\sum^{N}_{m=1} \alpha_m A^m.
\end{equation}

Equations (\ref{e_dotY}) and (\ref{e_intro_dotA}) form a differential equation
system whose initial conditions are $Y(0)=Y_0$ and $A(0)=A_0$. It was shown in \cite{Barberis2012} or, more rigorously, in \cite{Barberis2011}, that the components of the vector $Y$ can be written as:

\begin{equation}\label{e_Yi(t)}
 y_i(t)=\sum_{j,k=1}^n P_{ij}Q_{jk}y_{0k}f_N(\lambda_j;t),\,\,i=1..n,
\end{equation}
where $y_{0i} = y_{i}(0) $ is the initial condition, $P_{ij}$ and $Q_{ij}$ are the components of the matrices $P$ and $Q$ that diagonalize $A_0$ ($B_0 = PA_0Q$, with $QP = 1$ and $B_0$ diagonal), and the lowest order universal functions $f_N$ are given by,
\begin{eqnarray}
f_0(\lambda_j;t)&=&exp(\lambda_j t) \label{e_intro_fsub0}\\
f_1(\lambda_j;t)&=&exp \left[ \frac{\lambda_j}{\alpha_1}\left[ exp(\alpha_1 t)-1
\right] \right]\label{e_intro_fsub1} \\
f_2(\lambda_j;t)&=& \left\{ \frac{\alpha_2 \lambda_j}{\alpha_1} \left[
1-exp(\alpha_1 t) \right] +1 \right\}^{\frac{1}{\alpha_2}}.\label{e_intro_fsub2}
\end{eqnarray}

Here the truncation order $N$ determines the Vector Universality class VUN for a multiple-agent system, in analogy with the PUN classes developed by \cite{Castorina2006} for a single agent. The VUN elements are linear combinations of the functions $f_N(\lambda_j;t)$. Of course, the components $y_i(t)$ depend on the universality class, but we omit the index $N$ for simplicity.

\subsection{Interpretation.}

The solutions (\ref{e_intro_fsub0}), (\ref{e_intro_fsub1}), and (\ref{e_intro_fsub2}), which
depend on the eigenvalues $\lambda_j$ of the dynamical operator, correspond,
respectively, to joint Malthusian (VU0), joint Gompertzian (VU1), and joint
von Bertalanffian (VU2) growth processes. They apply to $n$ interacting agents
and are natural generalizations of the functions obtained in \cite{Castorina2006}
for the one-population problem.

Remarkably, the growth functions $y_i(t)$ depend only on the dynamic operator $A_0$, i.e., on $A_0(t)$ at a fixed time (the initial condition), which results in the uncoupling of the system for all subsequent times. This indicates that the interaction dynamics can always be characterized by the elements of $A_0$ at all times. The diagonal elements of this matrix determine the maximum size corresponding to the non-interacting populations in a given environment and thus its trace characterizes the global potential growth of the $n$ independent agents. We can say that each diagonal element defines the individual growth potential (IGP) of the corresponding agent in a given environment. Interactions among populations not only generate nonzero off-diagonal elements (the direct interactions), but may also modify the diagonal elements. This would occur if there were indirect influences among the populations, as it would be the case if the populations themselves modify the growth environment and, consequently, change 
the corresponding $A_0$ matrix element. We can thus describe various kinds of $n$-agent growth processes with $n^2$ real numbers, which allows us to quantify and classify various ecological-like interactions.

The information provided by the $A_0$ matrix after fitting can be summarized as follows: the direct interaction effects are represented by a combination of the signs and magnitudes of the off-diagonal matrix elements and determine how much an agent may gain (positive contribution) or lose (negative contribution) from the direct interaction. Then, for a parasitic interaction, one population grows at the expense of another (opposite signs), while for a cooperative (synergistic) interaction both agents benefit (positive signs) and for a mutual hindrance (antagonistic) interaction both agents are negatively affected (negative signs). The reciprocal of $\alpha_1$ defines a time scale that characterizes the joint growth rate.

Some representative  hypothetical  situations in the VU0 and VU1 classes are shown in the Appendix for the case of two populations. 

\section{Methods and Results} \label{s_results}

In this section we show how to use the VUN formalism as a tool to fit data from a cancer-related system with two  interacting cellular populations.

\subsection{Exponential growth}\label{s_pasquale}

Understanding the precise interplay of moving cells with other cells and their environment is crucial for central biological processes such as embryonic morphogenesis, wound healing, immune reactions, and tumor growth. As a  consequence, a large number of mathematical models, using multiple different approaches, have already been proposed to describe various aspects of cell motion  \citep{Keller1971,Murray1983,Dallon2001,Byrne2003,Dolak2005,Peruani2007,Hatzikirou2008}. In this section, we use experimental data to infer some features related to cancer cell migration, without any \emph{a priori} assumptions or models.

As we pointed out in the preceding section, the single-agent $N = 0$ case leads to simple exponential growth. However, already in the zeroth-order approximation VU0, which describes the growth of two or more interacting agents with infinite carrying capacity, new and more interesting situations arise: even if there is enough room and food for each individual agent to grow, VU0 describes how the interactions may lead to the stabilization and even the annihilation of one or more populations. In Appendix A we show some examples of how these situations can arise.

\begin{figure} 
\centering
\includegraphics[width=0.5\textwidth]{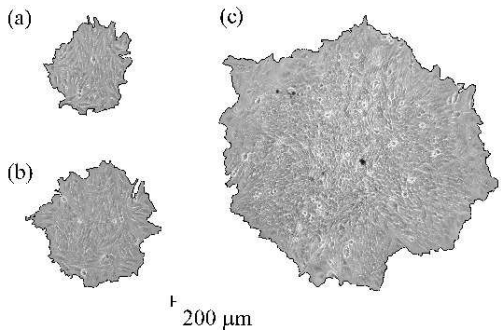}
\caption{Cell culture showing proliferative (bright rounded) and motile (elongated) cells at several times: (a) 5600 min. (b) 8300 min. (c) 15500 min.  } \label{f_HeLa}
\end{figure}

To illustrate VU0 growth in the cancer context, we use data sets obtained from HeLa cells in spreading colonies. These describe the correlated dynamic behavior of cells with different phenotypes: cells under cytokinesis (proliferative phenotype) or migrating cells (motile phenotype).To study the joint time evolution of those cells with different phenotypes, we made the follow-up of colonies from 4-10 cells up to about 1500 cells over 8-12 days. HeLa cells, passage 44-60 (a gift from Leloir Institute), were seeded into polystyrene Petri dishes (3.6 cm in diameter) from 2 ml of RPMI growth medium containing 1000-2000 cells/ml and incubated in an oven at 37 C in 5 \%  carbon dioxide and 97 \% humidity. Cells were allowed to adhere and grow for 4-5 days forming small colonies of 4-10 cells. To maintain unchanged environmental conditions, half of the medium was changed every two days. Colony patterns were followed by optical microscopy by taking pictures every 15 hs. All cells in colonies either under cytokinesis 
or moving were counted. The former look bright and rounded and the latter have rather elongated shape as shown in Fig. \ref{f_HeLa}. Eventually cell trajectories that characterize their motion were obtained from \emph{in situ} cell colony digital images acquired by a time-lapse system at intervals in the range of 5-45 min. during 2–3 days. For this purpose, colonies were placed inside a chamber that was fixed to the microscope platform, and maintained at 37 ºC and 97 \% humidity. In order to preserve the pH, the culture medium was changed to L-glutamine supplemented RPMI CO2-Independent Medium (Gibco, In-vitrogen Corp.) before placing each Petri dish into the chamber.

In Fig. 2a, the resulting counts for both populations are shown together with a possible VU0 fitting. Since the optimization problem has several local minima, care must be taken in the choice of the initial conditions for the fitting parameters. For instance, if we look at a restricted time interval, we could be tempted to accept the fitting in the figure, which really corresponds to a ``wrong'' (W) set of parameters. In this case, the fit would tell us that the initial population for the motile phenotype ($y_{01} = 20 $) cells is twice the experimental one, and that there are also some proliferative cells at $t = 0$ ($y_{02} > 0$). Extrapolating to long times, W would predict an unlikely rapid decay of the motile population, as shown in Fig. 2b. Since the resulting $A_0$ matrix predicts a parasite-like interaction in favor of the proliferative population, the motile population cannot survive at very long times. On the other hand, a set of ``correct'' (C) parameters can be found that provides the fitting 
shown in Fig. 3a, which looks very similar to 
that in Fig. 2a but gives an adequate prediction for the initial populations (within a reasonable error). As it can be observed in Fig. 3b, the set C leads to a  plausibly unrestricted growth of both populations at long times.

The set of parameters for both fittings is given in detail in Table I. Note that the $R^2$ statistical parameter is not a suitable criterion to discriminate between the C and W fits. In situations where there are no physical clues about the right parameter set, we must use more reliable criteria, such as Akaike's (Burnham and Anderson, 2002) or the more complex Maximal Information Coefficient from Reshef et al. (2011).

\begin{table}
 \begin{tabular}{l|rrrr}
  \toprule 
 &$ a_{11}$& $a_{22}$&$ a_{12}$& $a_{21}$ \\ 
Correct&0.000548&-0.000049&-0.008105&0.000011 \\ 
Wrong&0.000459&0.000346&-0.004848&0.000001 \\\midrule
  &$y_{01}$&$ y_{02}$&$ R^2_{1}$&$ R^2_{1}$\\
Correct&9.90&-0.20&0.9918&0.9810\\
Wrong&20.19&0.40&0.9917&0.9855\\\bottomrule
 \end{tabular}
\caption{Parameters obtained using VU0 fitting for the ``correct'' and ``wrong'' cases. See text.}\label{t_exp}
\end{table}

The $A_0$ matrix obtained from the C fitting has $a_{12} < 0$, meaning that the number of cells with the motile phenotype is decreased by their interaction with the proliferative cells. Since motile cells can only replicate  by changing their phenotype to proliferative, replicating and then changing again to motile, the motile population loses some cells in order to increase their number later. This decrease in the motile phenotype  population corresponds to the temporary loss of those cells that stop for replication. On the other hand, $a_{21}\gtrsim0$, representing the increase in the proliferative population due to the transformation of motile cells.

The diagonal elements have opposite signs favoring the motile phenotype. The motile phenotype has $a_{11}>0$,  which tells us that there is a net positive feedback of its transformation into the reproductive phenotype. The net environmental conditions favor the transformation of reproductive cells into the energy-saving motile phenotype, making its population much bigger. The resulting  population difference between phenotypes is in 

The negative sign of $a_{12}$ suggests the presence of a signal originating in the proliferative cells that tells their motile counterparts that they should stop and reproduce. Cell behavior in this system reminds us of the ``go or grow'' hypothesis sometimes used to explain the onset of metastasis in gliomas and other cancers \citep{Corcoran2003,Hatzikirou2012,Gerlee2012,Garay2013}. This hypothesis postulates that migration and proliferation are spatiotemporally mutually exclusive.

\begin{figure} 
\centering
 \subfloat[][]{
\includegraphics[width=0.45 \textwidth]{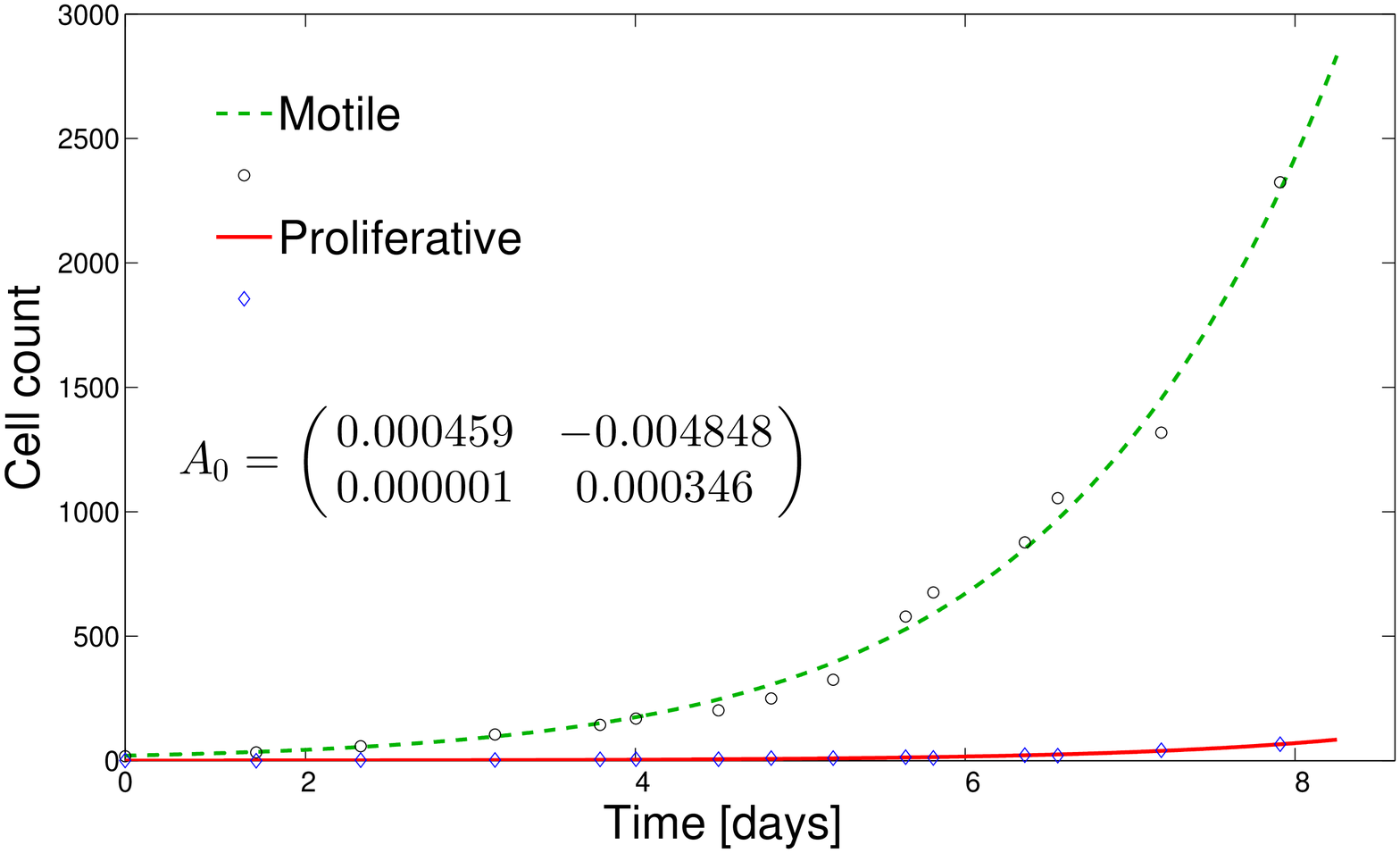}
\label{f_HeLa_M_VU0}
}
\qquad
\subfloat[][]{\includegraphics[width=0.45 \textwidth]{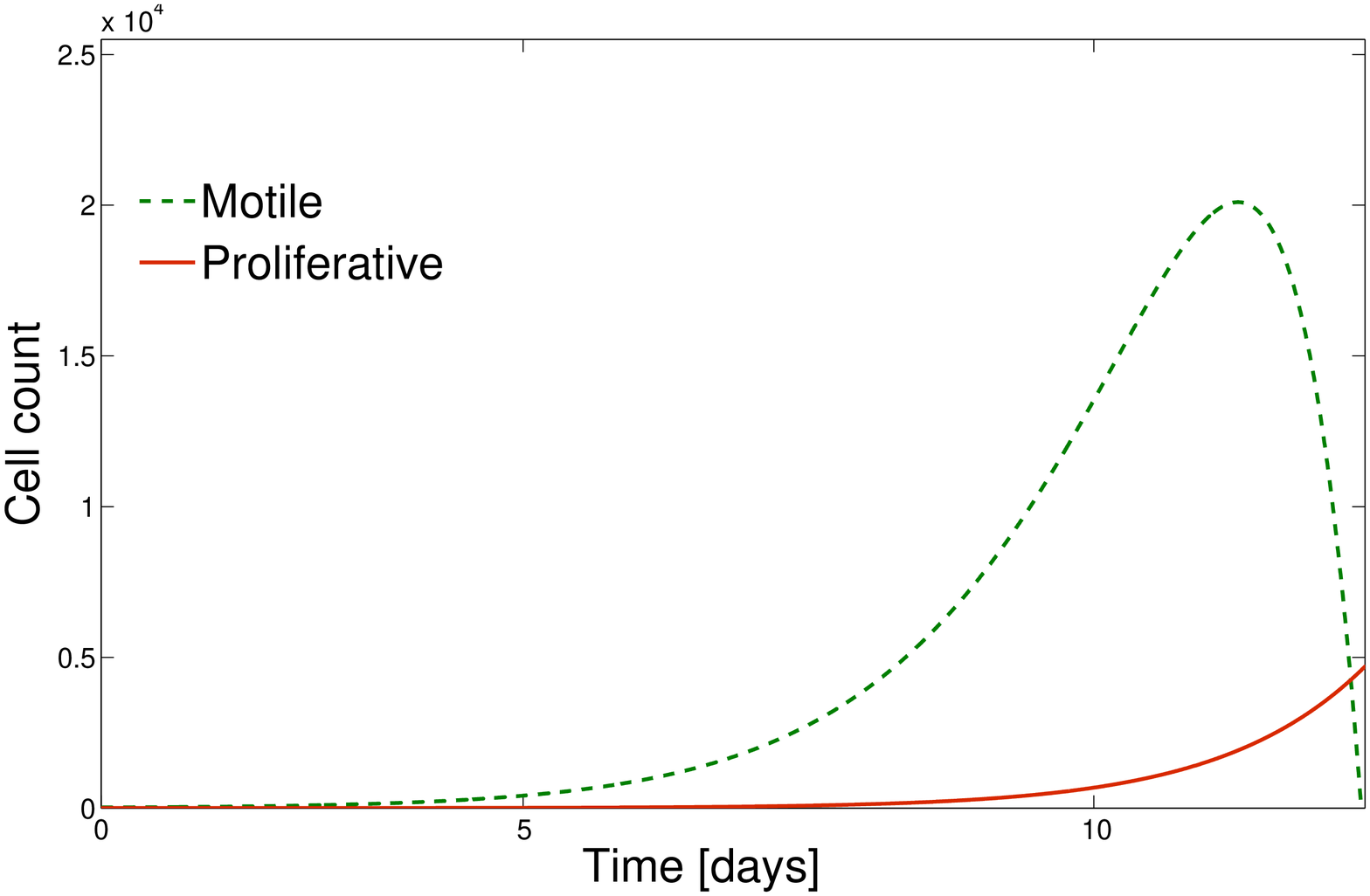}
\label{f_HeLa_M_Pred}
}
\caption{(a) Possible VU0 fitting of the population data for motile and proliferative HeLa cells. (b) Changing the scales we observe that the resulting parameters are inadequate. Care must be taken when choosing the local minimum. } \label{f_HeLa_M}
\end{figure}

\begin{figure} 
\centering
 \subfloat[][VU0 fitting.]{
\includegraphics[width=0.45 \textwidth]{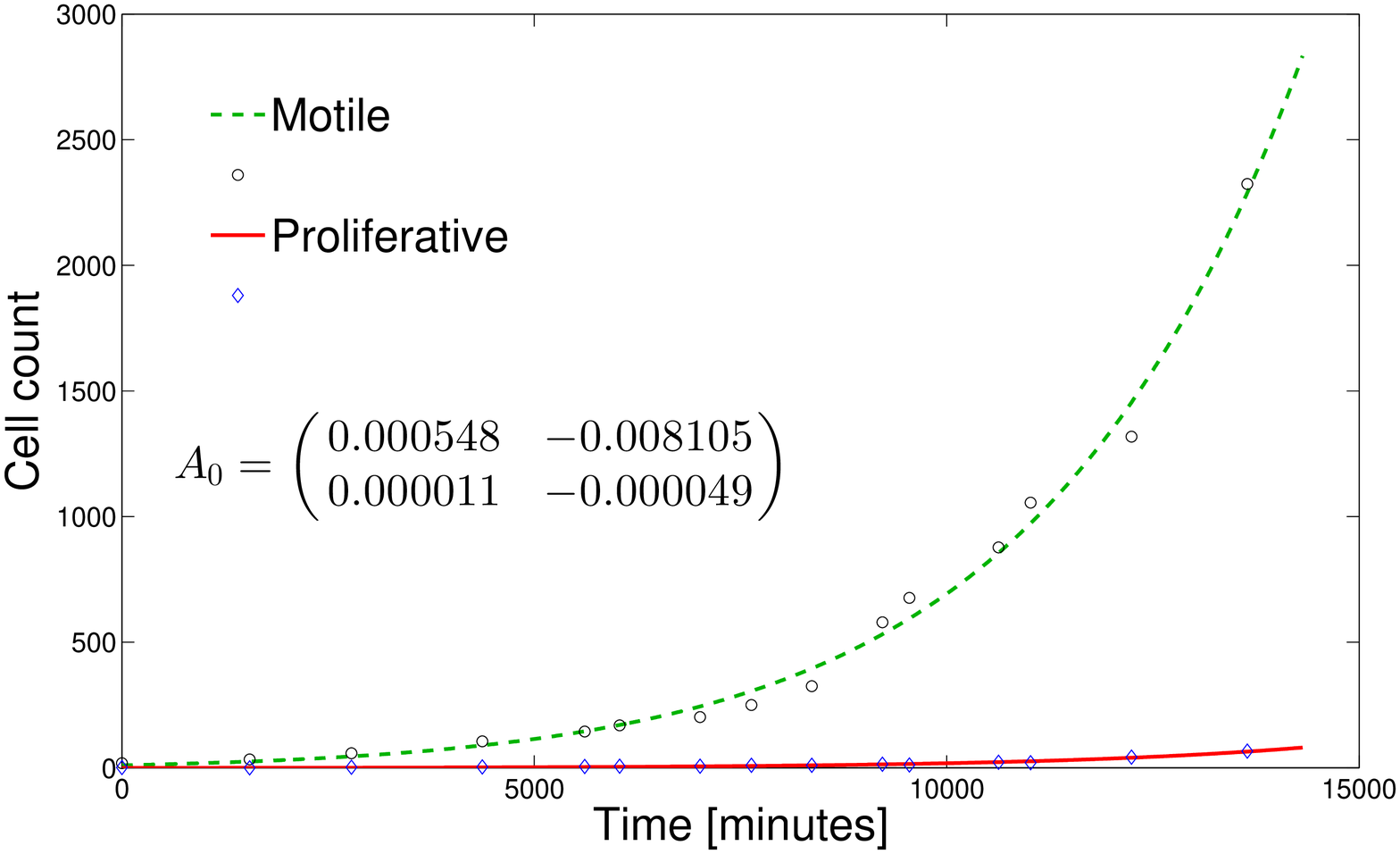}
\label{f_HeLa_B_VU0}
}
\qquad
\subfloat[][Good prediction]{\includegraphics[width=0.45 \textwidth]{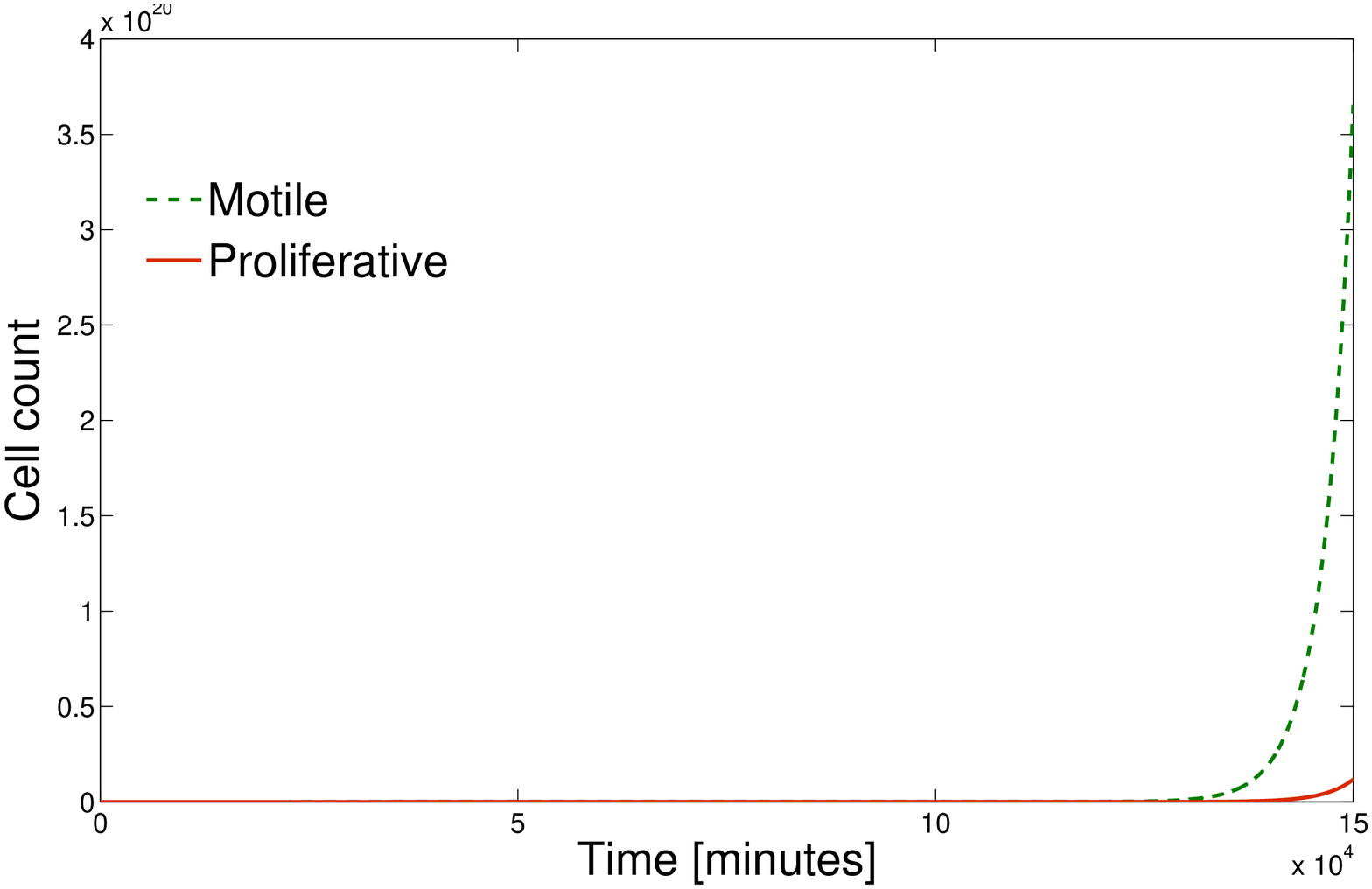}
\label{f_HeLa_B_Pred}
}
\caption{ (a) VU0 fitting of the population data for motile and proliferative HeLa cells. (b) The rescaled figure suggests that the parameter set is adequate. The $A_0$ matrix allows us to infer that the motile cells are not able to replicate themselves.   } \label{f_HeLa_B}
\end{figure}

\subsection{Multicellular Tumor Spheroids growth}\label{s_MTS}

\begin{figure} 
\centering
 \includegraphics[height=5 cm]{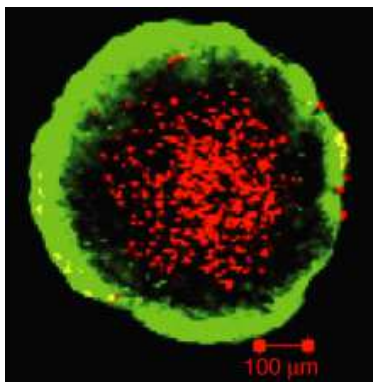}                                                                                                                                  
\caption{A multicellular tumor spheroid (MTS) where the live cells in the outer rim appear as a green halo. The red mass inside belongs to the necrotic core.\dag  } \label{f_MTS}
\end{figure}

Multicellular Tumor Spheroids (MTS) are spherical clusters of tumor cells that may be grown \emph{in vitro} under strictly controlled conditions (Hamilton, 1998; Mueller-Klieser, 2000). MTS experiments have led to new insights in cancer research. In particular, they have been useful to characterize the dependence of necrosis formation on the external environment (Freyer and Sutherland, 1988). During spheroid growth the fraction of proliferating cells decreases and the cells in the inner region become deprived of oxygen, glucose and other nutrients whereas metabolic waste accumulates, and the formation of a necrotic core is observed. In an advanced growth stage spheroids exhibit  an outer viable rim (whose thickness ranges from about 100 to 250 $\mu$m) that surrounds the necrotic core. The spheroid eventually attains a limiting size with a final diameter of $1$ - $3$ mm that \cite{Folkman1973}, attributed to the degradation of dead cells in the necrotic core \cite{Bertuzzi2010}. An example of MTS is shown in 
Fig. \ref{f_MTS}, where the live cells in the outer rim appear as a green halo. The red mass inside belongs to the necrotic core. 

The main advantages of the MTS are their simple spherical symmetry and the fact that they can be produced in large quantities. These characteristics make them popular both in mathematical modeling and  in biological research: MTS provide an experimental biological model that has allowed researchers to determine protein expression \cite{McMahon2012,Gupta2012}, check mathematical models \cite{Radszuweit2009,Bertuzzi2010,Kazmi2012}, and study drug delivery to treat cancer disease \cite{Mehta2012,Gibot2013}, to mention a few applications. The understanding of how MTS grow is indeed crucial to further comprehend  some aspects of \emph{in vivo} tumor progression.

We may start our study with the hypothesis postulated by Guiot and collaborators \cite{Guiot2003,Delsanto2004}, who demonstrate that the growth of homogeneous spheroids can be fitted by the von Bertalanffy-West equation. Such a description was later generalized to spheroids with a necrotic core by \cite{Condat2006} and \cite{Menchon2007}, who added the proposal that the time evolution of the total mass $M$ of the spheroid, as function of its live cell mass $m_v$, can be written as:

\begin{equation}\label{e_necro}
 \frac{dM}{dt}=\frac{a\tilde{m}^{2/3}}{12}\left[ \left(
4\frac{m_v}{\tilde{m}}-1\right)^{1/2}+ \sqrt{3} \right]^2-bm_v;
\end{equation}
where  $\tilde{m}$ is the mass at the onset to necrosis, $a$ and $b$ are the metabolic rates defined by \cite{West2001} and the exponent $2/3$ describes the diffusion-limited energy influx to the system via the transport of  glucose and oxygen. 
The key to obtain Eq. (\ref{e_necro}) is the invariability of the thickness $\Delta$ of the outer rim of live cells. This feature of the MTS was first described by \cite{Burton1966}, and later discussed by \cite{Groebe1996}.
With the aim to go beyond these works, we add to equation (\ref{e_necro}) the time evolution of the dead mass $m_m=\frac{4}{3}\pi \rho [r(t) - \Delta(r(t))]^3$ as  a function of the radius $r$ of the whole spheroid. Combining it with the onset-to-necrosis mass $\tilde{m}=\frac{4}{3}\pi \Delta^3$, we obtain the equation set:

\begin{subequations}
\label{e_necroticoLucas}
\begin{align}
  \frac{dM}{dt}&=a M^{\frac{2}{3}}-b \left[ M -
(M^{\frac{1}{3}}-\tilde{m}^{\frac{1}{3}})^3\right],\label{e_total}\\
 \frac{dm_{m}}{dt}&=
M^{-\frac{2}{3}}(M^{\frac{1}{3}}-\tilde{m}^{\frac{1}{3}})^2
\frac{dM}{dt}, \label{e_muerta}\\
 \frac{dm_{v}}{dt}&=\frac{dM}{dt}- \frac{dm_{m}}{dt},\label{e_viva}
\end{align}
\end{subequations}

which describes the time evolution of the dead and live cells populations \emph{after} the onset of necrosis.

\begin{figure} 
\subfloat[][]{
 \includegraphics[height=5 cm]{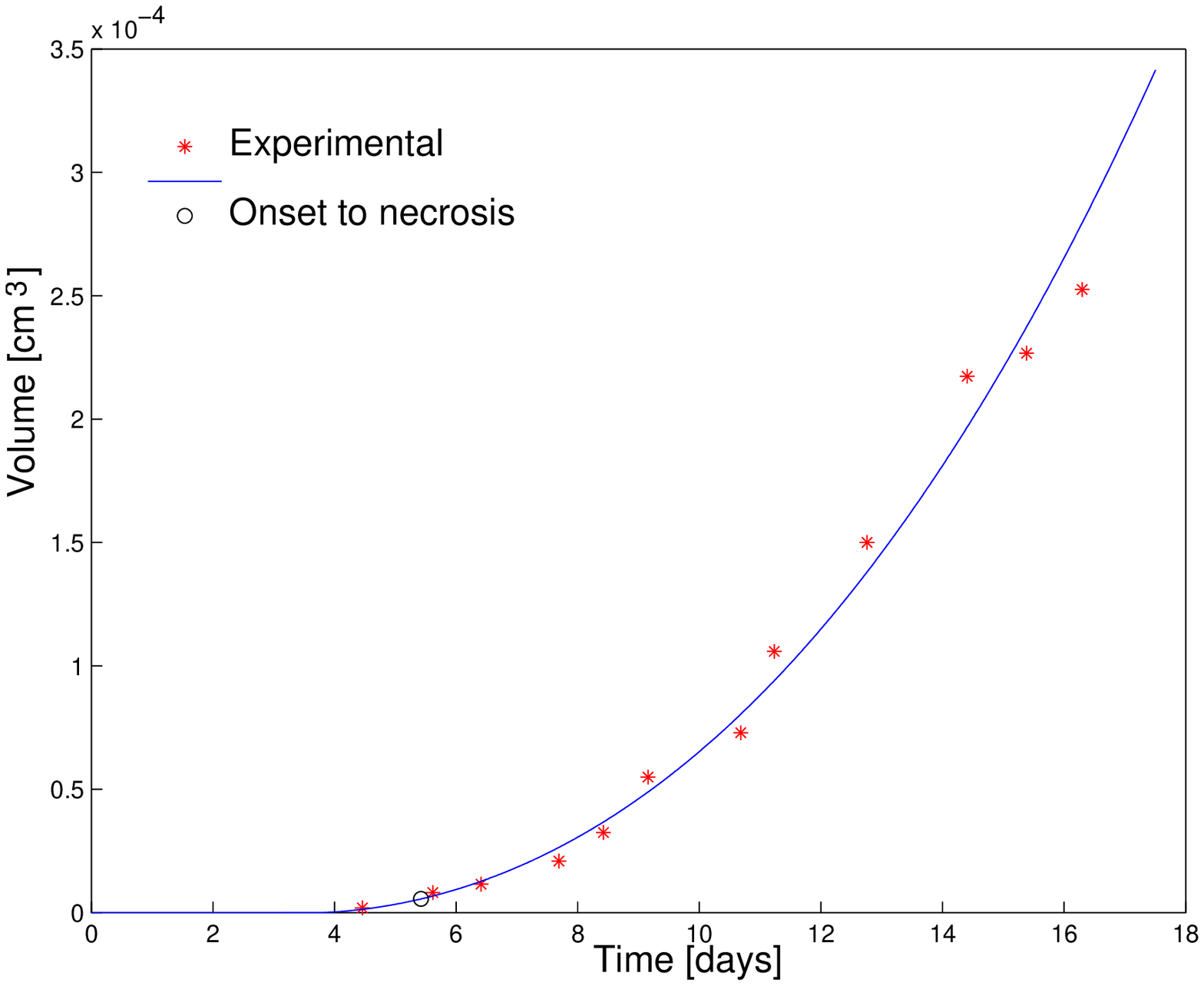}\label{f_MTS_recon_fit}
}
\subfloat[][]{
\includegraphics[height=5 cm]{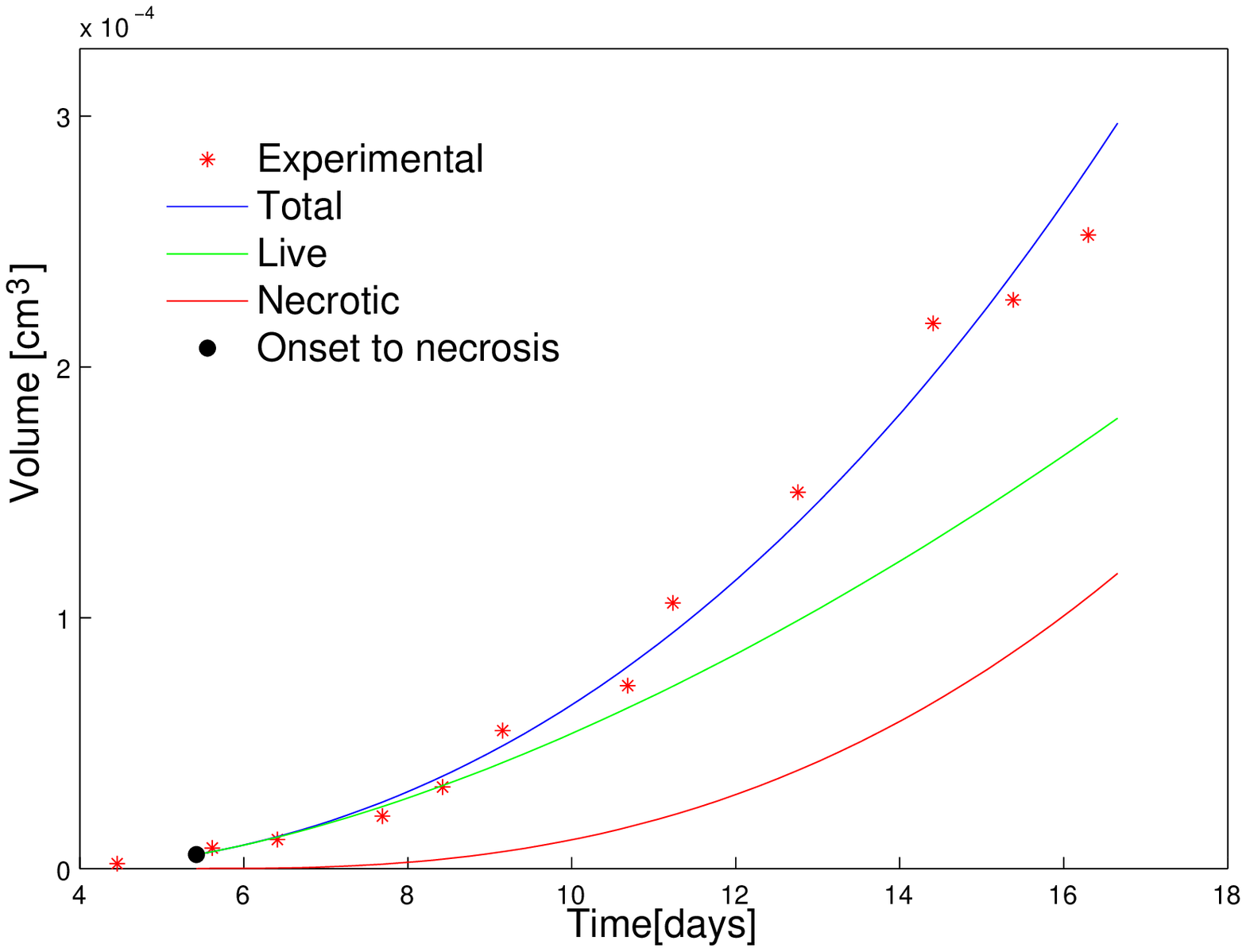}\label{f_MTS_recon_build}
}
\caption{Cellular populations reconstruction. (a) Fitting of the experimental data by means of Eq. (\ref{e_total}). (b) Predicted populations by means of Eqs. (\ref{e_muerta}) and (\ref{e_viva}).  }\label{f_MTS_recon}
\end{figure}

\subsubsection{In vitro MTS.}\label{ss_vitro}                                                                                                                                                                                                                                                                                                                                                                                                                                                                                                                     

First we consider the experimental data of \cite{Freyer1986}, who created large numbers of MTS using EMT6/Ro cancer cells. They grew them at several oxygen and glucose concentrations in order to measure spheroid growth rates and the thickness $\Delta$ of the outer rim  as functions of MTS diameter. We will use the VU1 equations to ascertain the individual time dependences of the necrotic and live masses, which were not separately measured in the experiment. To do this, we first obtain the parameters $a$ and $b$, using the value of $\tilde{m}$ reported by  \cite{Freyer1986}  to fit their data with the numerical solution of Eq. (\ref{e_total}), as shown in Fig. \ref{f_MTS_recon_fit}. We remark that not all of the data series reported in \cite{Freyer1986} can be adequately fitted because Eqs. (11) are only valid after the onset of necrosis. In those experiments in which the necrosis starts later, there are not enough data points to provide an acceptable fit. The example 
presented here 
corresponds to the experiment with $16.5$ mM of glucose and $0.07$ mM of oxygen. Figure \ref{f_MTS_recon_build} shows the reconstruction of the time evolution of the populations by numerical integration of Eqs. (\ref{e_viva}) and (\ref{e_muerta}). The fit yields the following values: $a = 0.0089$ g$^{1/3}$/day; $b = 0.0097 1$ 1/day, and $\tilde{m} = 5.57$ $\mu$g. With these values of the parameters, the solutions of Eqs. (11) describe unrestricted growth at long times, which is not observed in the experiments.  

If we observe Fig. \ref{f_MTS_recon_fit}, we note that, at first sight, the data points suggest  exponential growth. Freyer and Sutherland ascribed such rapid growth to the initial part of a Gompertz curve. To make the saturation visible, they presented the data using a semilog graph. This saturation of growth is always observed in an MTS, even if the cell cumuli are placed in a culture media with enough room and nutrients to grow indefinitely. The authors solved this apparent paradox by suggesting that the necrotic cells secret growth inhibitor substances that affect the live cells.  More than a decade afterwards, \cite{Wartenberg1999}, among others, experimentally confirmed this hypothesis.   Because the action of such inhibitors is not taken into account in the deduction of Eqs. (11), their long-time predictions may not be accurate. But in the early development of the MTS described in the experiment, for which the  necrotic core size is small, its inhibitory action 
is weak enough to allow for reliable data fitting with the solutions of Eqs. (\ref{e_necroticoLucas}). However, this weak inhibitory effect suffices to eventually lead to saturation, an effect that is well described when VU1 is used to simultaneously fit both populations, as we show next.  

\begin{figure}
\subfloat[][]{
 \includegraphics[width=0.5 \textwidth]{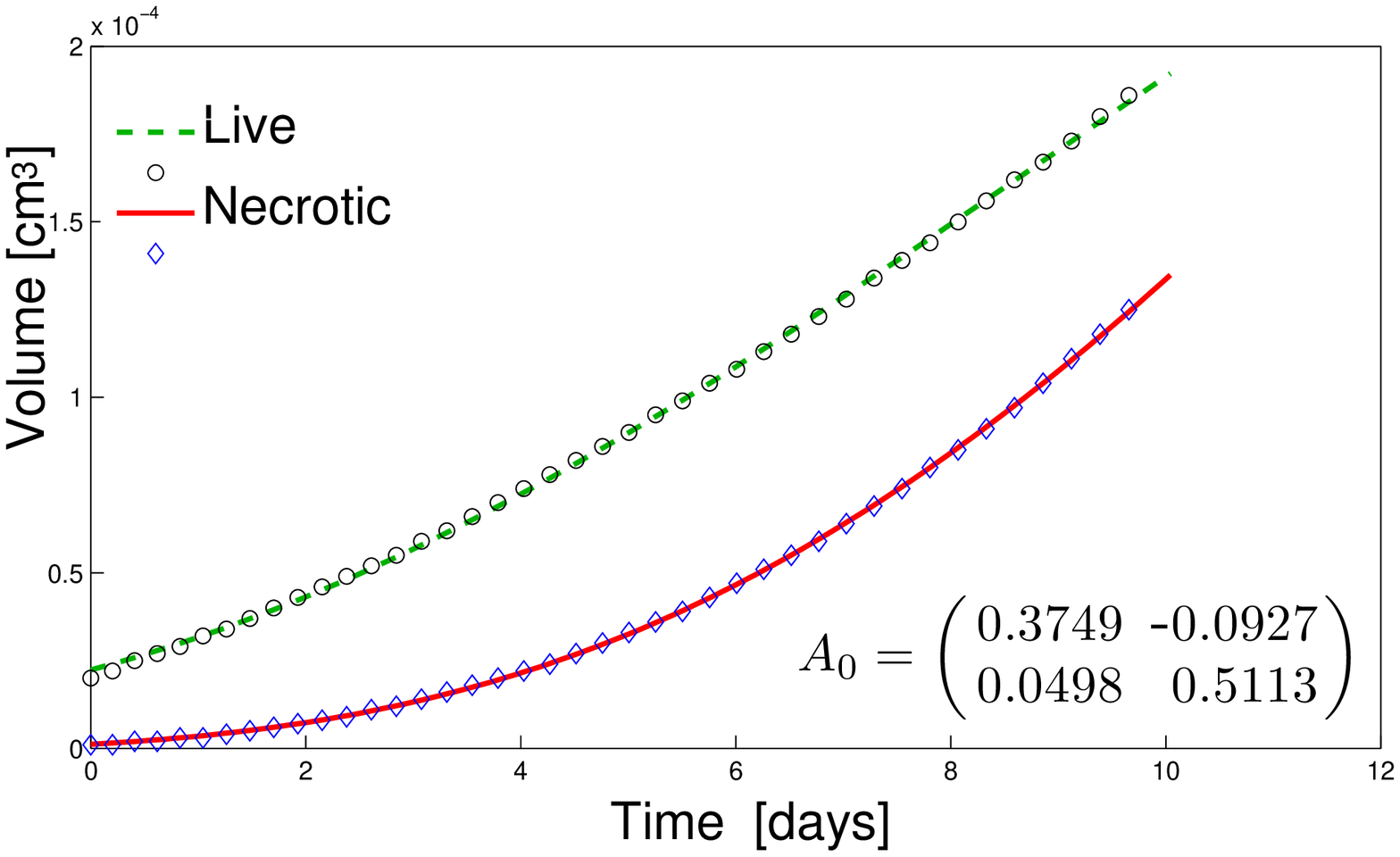}\label{f_MTS_freyer_fit}
}
\subfloat[][]{
\includegraphics[width=0.5 \textwidth]{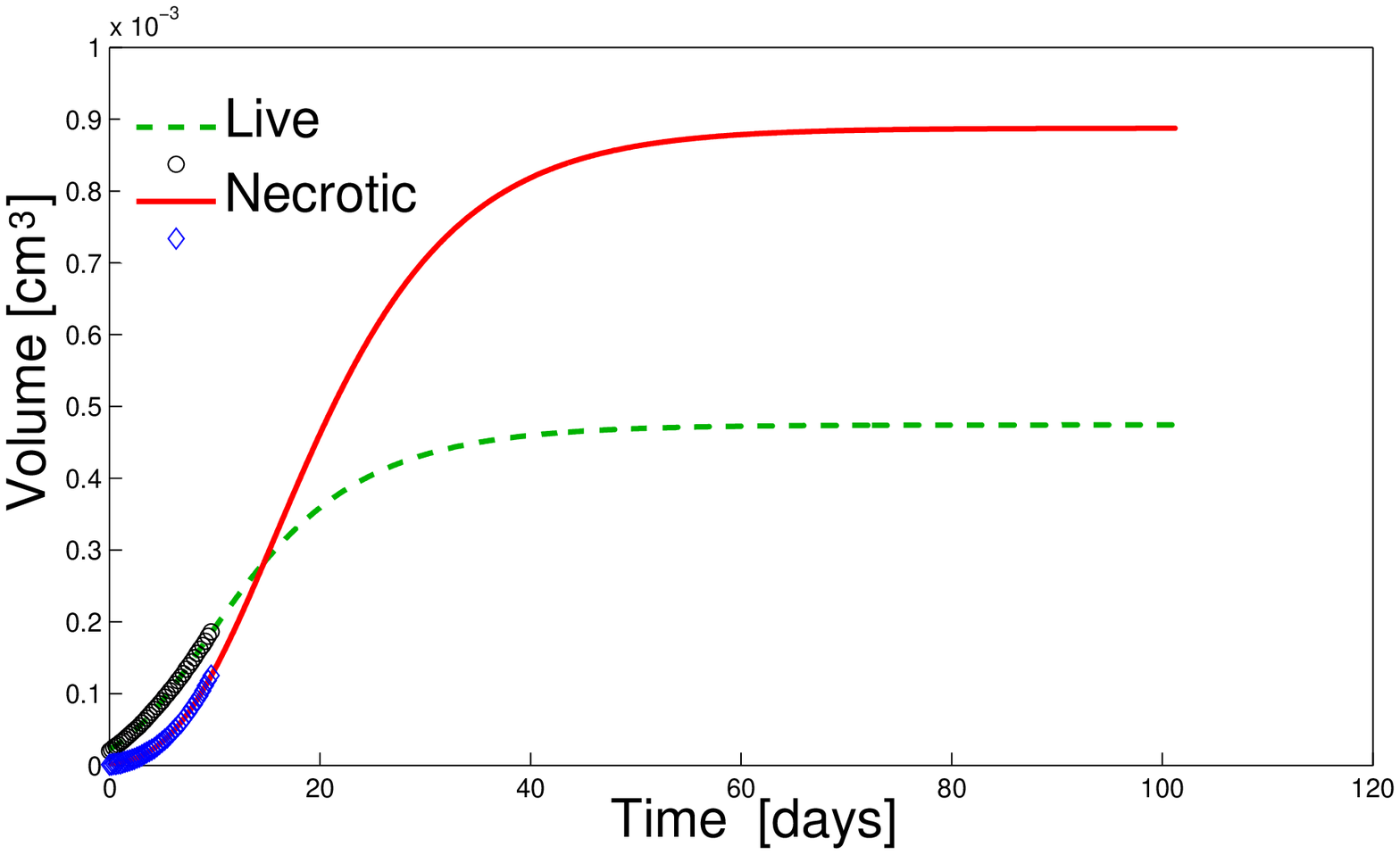}\label{f_MTS_freyer_long}
}
\caption{ VU1 fitting for the reconstructed  live and necrotic population datasets for  \emph{in vitro} spheroids.
(a) Experimental data range. The $A_0$ matrix predicts the inhibition ($a_{12}<0$) of the live cells by the dead ones.  
(b) Extending the growth functions beyond the time of the data range, VU1 is able to predict an asymptotic volume $V_{\infty}\simeq 1.3 \times 10^{-3}$cm$^3$, which is reasonable for this system.
\label{f_freyerVUN} }\label{f_MTS_freyer}
\end{figure}

After using Eq. (\ref{e_total}) to obtain parameters $a$ and $b$, synthetic data sets for the live and dead cell populations were obtained by numerical integration of Eqs. (\ref{e_muerta}) and (\ref{e_viva}). We then used VU1 to simultaneously fit the time dependences of the two cell populations. The outcome of the fit, which is shown in Fig. \ref{f_MTS_freyer_fit}, is remarkably informative. First, since $a_{12} < 0$, the $A_0$ matrix predicts an inhibitory process of the growth of live cells due to the presence of a necrotic population. The necrotic population growth is favored by the live cells ($a_{21} > 0$) because the death of the latter is the source of the former. The meaning of the diagonal terms is straightforward: $a_{11} > 0$ because live cells generate more live cells and $a_{22} > 0$ because dead cells generate more dead cells. Thus the VUN fitting provides further proof of the presence of an agent that is generated by dead cells and acts as both a growth inhibitor and a facilitator of cell 
death.

Furthermore, VU1 fitting predicts the saturation of the whole population (equations (11) do not). The predicted asymptotic volume is $V_{\infty}\simeq 1.3 \times 10^{-3}$cm$^3$. This value is three times lower than the one obtained by \cite{Mueller2000}  for well-nourished tumors ($ 4 \times 10^{-3}$cm$^3$). The discrepancy presumably arises because the glucose concentration used in the experiment is half what  Freyer and Sutherland consider optimal: $0.8$ mM instead of $16.5$ mM. We remark that the oxygen level ($0.28$ mM), which  is more essential, is at its optimum value. The predicted required time to reach the stationary regime, near $40$ days, is much longer than the time explored in the experiment ($\simeq 28$ days). But as the time scale from the fitting is $\frac{1}{\alpha_1} \simeq 9.7$  days, which belongs to the data range, we may accurately infer the asymptotic value. This inference is possible because the stabilization is mainly  caused by the parasite-like 
interaction between necrotic and live cells.

Finally, we tried to fit the data with VU2. These attempts were unsuccessful.  The impossibility of finding a find a suitable fitting suggests that this system should be classified as VU1.  As far as we know, this is arguably the first proven  truly Gompertzian system found in the PUN literature  \citep{Delsanto2008a,Pugno2008,Gliozzi2010,Gliozzi2011,Gliozzi2012,Mazzetti2012}. These authors have always found   von Bertalanffian  systems.
Furthermore, a few years ago, \cite{Gonzalez2006} proposed that the Gompertzian function can be obtained from a differential system that involves two populations, one alive and another quiescent. This is exactly the case of an MTS with necrotic core. So the VU1 fit recovers the dynamics of a system of live and quiescent populations without any \emph{a priori} assumptions on the nature of the data.

\subsubsection{In vivo MTS.}\label{ss_vivo}

Next we apply the VUN formalism to implanted \emph{in vivo} tumors. The data about tumor progression was taken from \cite{Steel77}. In this case,  MTS of the khjj cell line  were grown \emph{in vitro} and then implanted in mice. Although data on the necrosis is not available,  \cite{Menchon2007} used Eq. (\ref{e_necro}) to show that the addition of a necrotic core leads to a very good fit to the experimental data, while a fit that assumes a homogeneous population fails.
Implementation of  Eqs. (11) yields estimates that are not too different from those of \cite{Menchon2007}.  Note that, in this work, the mass  at the onset to necrosis is also taken as a fitting parameter. The parameter values we obtain are $\tilde{m} = 0.03496$ g, $c= 0.2812$ g$^{1/3}$/day and $b =0.3515$ 1/day.

\begin{figure} 
\subfloat[][]{
 \includegraphics[width=0.5 \textwidth]{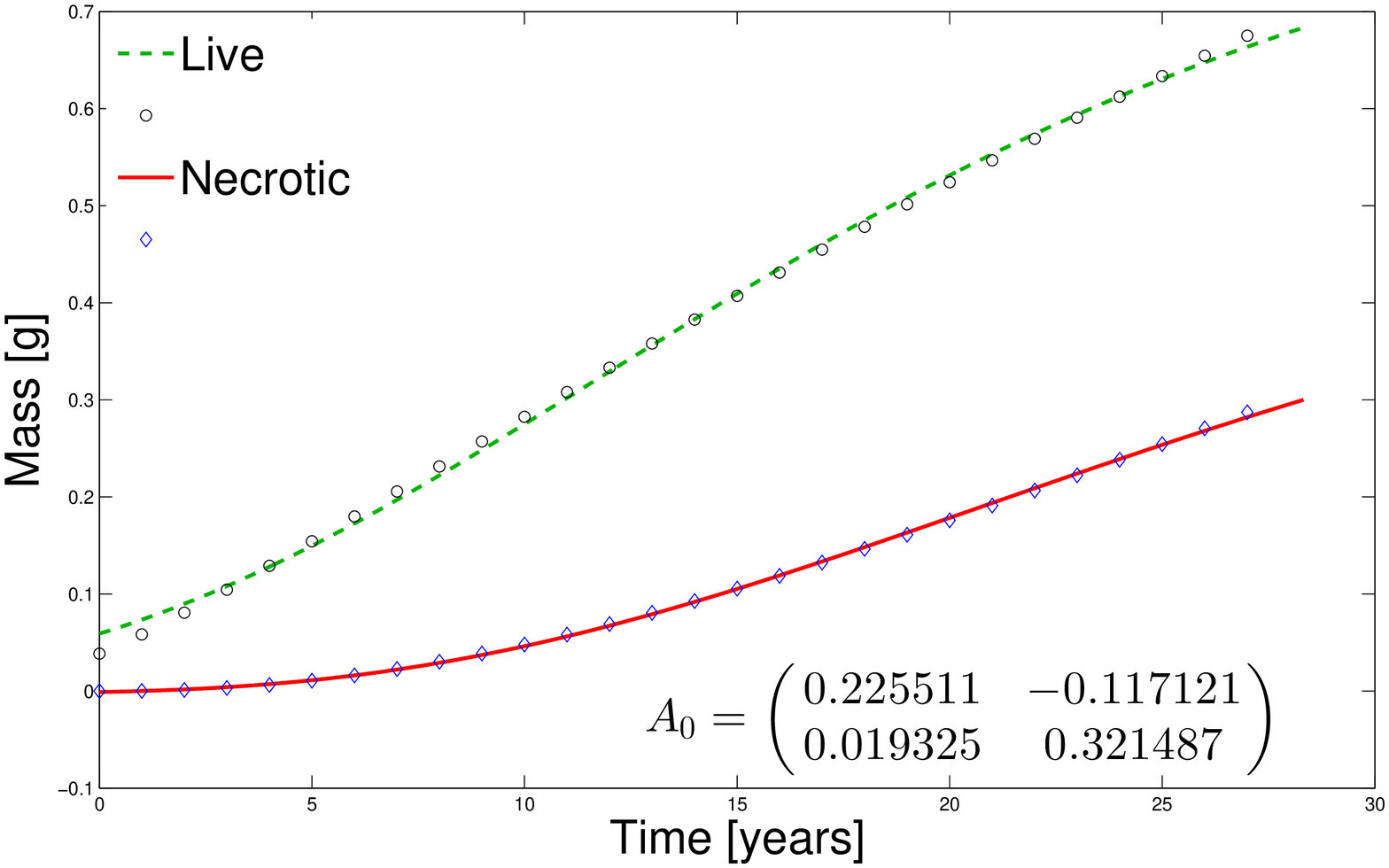}\label{f_MTS_steel_1}
}
\subfloat[][]{
\includegraphics[width=0.5 \textwidth]{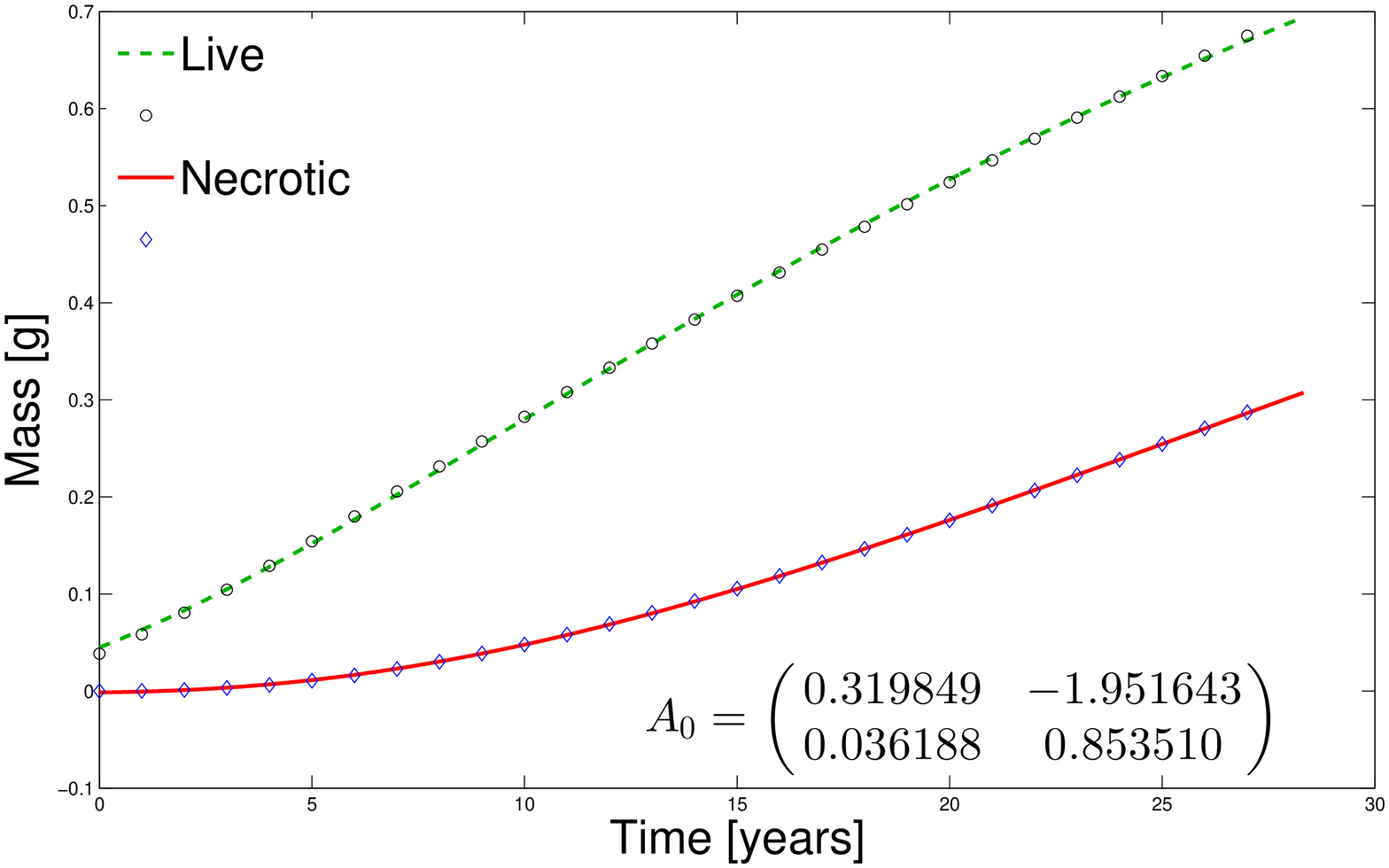}\label{f_MTS_steel_2}
}
\caption{ Fitting of the reconstructed \emph{in vivo}  data set corresponding to khjj spheroids (Steel 1977). (a) VU1 fit: The $A_0$ matrix components are of the same order but a little lower than in the \emph{in vivo} case. (b) VU2 fit, which is more accurate than its VU1 counterpart. 
 }\label{f_MTS_steel}
\end{figure}

In Fig. \ref{f_MTS_steel_1}, we present a VU1 fitting of the reconstructed data for this \emph{in vivo} tumor. The values for the determination coefficients are not as remarkable as those  for the \emph{in vitro} case discussed above but they are still very good: $R^2_1=0.99936$ and $R^2_2=0.99931$. Of course, implanted tumors lose their spherical shape at long times and the equations (11) become a less accurate approximation. The values for the $A_0$ matrix give results comparable to those in the \emph{in vitro}  experiment: the necrotic cells inhibit the growth of the live cells ($a_{12}<0$) and benefit from them ($a_{21}>0$) because  when the latter die they increase the necrotic population. 

The diagonal elements of $A_0$ corresponding to the data of Steel (1977) and of Freyer and Sutherland (1986) are not too different. This is noteworthy because they correspond to different cell lines and very different growth conditions. Since the necrotic cells do not compete among themselves but $a_{22}$ is reduced by a factor of about $2$, we may infer that this reduction is due to the different nature of the \emph{in vivo} experiment. Not only is the hydrostatic pressure in the \emph{in vivo} system greater than that in the \emph{in vitro} case, hindering cell multiplication and thus decreasing the value of $a_{11}$,  but the biochemical environment may be conducive to a faster removal of necrotic cells, leading to a decrease in the value of $a_{22}$, too.

The characteristic time  $\frac{1}{\alpha_1}$ is now significantly longer than in the \emph{in vitro} experiment (over 13 days against 9.7 days). This difference  can be explained by the hostility of the \emph{in vivo} environment (expressed through the enhanced hydrostatic pressure and the immune response of the host organism), which leads to a slower growth of the tumor implanted in a mammal host as compared to that of an \emph{in vitro} system. 

Here we also tried a VU2 fit, which is shown in Fig. 7b. The fit turns out to be of higher quality than that obtained from VU1, as we can visually recognize (note, for instance, the improved accuracy at the ends of the live cell curve), and confirm from the higher values obtained for $R^2$. The accuracy of the VU2 fit (inexistent for the spheroids of 3.2.1) seems to indicate the higher complexity of the \emph{in vivo} system, which may demand a more complex class of growth functions. All the $a_{ij}$ matrix elements preserve their signs and the interpretation stemming from the VU1 study, but their magnitudes are magnified. The joint characteristic time for VU2 is $\frac{\alpha_2}{\alpha_1}\simeq 9$   days, shorter than its VU1 counterpart. Here we fixed the value  $\alpha_2=-1/3$ in order to describe diffusion-limited feeding \cite{West2001,Delsanto2004}. Table \ref{t_MTS} exhibits the parameter values we found for this problem.

In summary, the above \emph{in vitro} and \emph{in vivo} examples show that, at least in their initial stages, tumors growing in very different environments exhibit similar macroscopic behavior, with the number of live cells initially growing very fast and then saturating, while a necrotic core emerges, which is considerably smaller \emph{in vivo}. Both cases are well described by the VU1 universality class, although VU2 may be preferred for the more complex \emph{in vivo} systems.

\begin{table}
 \begin{tabular}{c|ccccc}
  \toprule 
 &$ a_{11}$& $a_{22}$&$ a_{12}$& $a_{21}$&$\alpha_1$\\\midrule 
\emph{In vitro}&  &&&&   \\
VU1&0.3749&0.5113&-0.0927&0.0498&-0.104 \\\midrule 
\emph{In vivo} &&&&&\\
VU1&0.2255&0.3215&-0.1171&0.0193&-0.075 \\
VU2  &0.3198&0.8535&-1.9516&0.0362&-0.038\\\bottomrule
 \end{tabular}
\caption{Parameters obtained by fitting the reconstructed MTS subpopulations.}\label{t_MTS}
\end{table}

\subsection{Growth of Subspheroids}\label{s_subspherois}

Cell shedding by a primary tumor is an important factor in the formation of metastatic colonies. It is controlled by adhesiveness \cite{Menchon2009}. Adhesiveness can be increased by polyphenols, which may then inhibit cell escape \cite{Gunther2007}. In an experimental study of invasion and metastasis, G\"unther et al. (2007) grew mammalian MTS of the breast cancer cell line 4T1. The aim of their work was to understand the factors that regulate the shedding of cells from primary tumors as functions of time and the size of the subspheroids that grow from the detached cells. 

The panels in Fig. 8a, taken from  G\"unther (2007), show the spheroid populations on days 4, 6 and 9 of culture, respectively. In the right panel we observe the subspheroids grown from the cells detached from the primary spheroids. In this case it is not possible to find a suitable VU2 fit, but the VU1 fit is completely satisfactory. As shown in Fig. 8b, even without assuming the absence of subspheroids at $t = 0$, it is possible to obtain a fit able to predict the absence of a detectable subspheroid population for over three days. The diagonal elements of the $A_0$ matrix are very different to account for the rather unequal characteristic growth times. Indeed, very different eigenvalues of $A_0$ are needed in order to generate two growth functions with so dissimilar growth rates. 

The high value of $a_{12}$ suggests that the primary spheroids derive a large benefit from their interaction with the subspheroids. The detachment of peripheral cells allows more central ones to get nutrients, stay alive and even reproduce. On the other hand, the detached cells must survive competing for nutrients and space with the primary spheroids, which, at this stage, are hard contenders. Therefore, it is not surprising that $a_{21} < 0$, even if its magnitude is very small. But subspheroid growth is strong because (a) they derive from very active shed cells, and (b) their initial small size allows them to occupy spaces with room to grow and high nutrient concentration. The high value of $a_{22}$ can then be understood as related to the enhancement of the survival capability of the shed cells because they have passed a natural selection test.

\begin{figure} 
\subfloat[][]{
 \includegraphics[width=\textwidth]{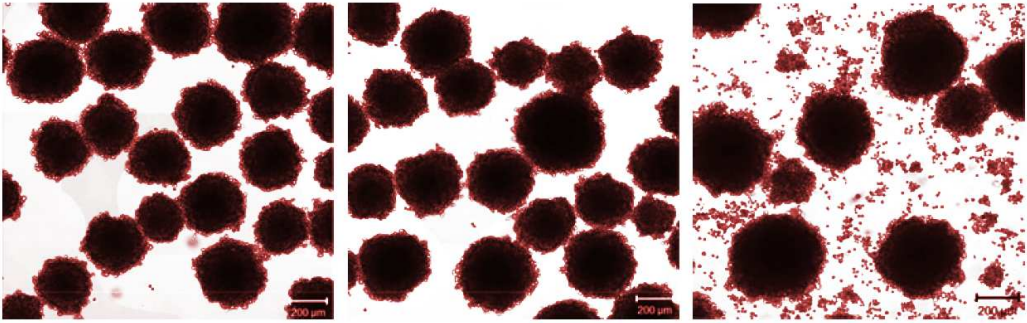}\label{f_sub_fig}
}\\
\subfloat[][]{
\includegraphics[width=\textwidth]{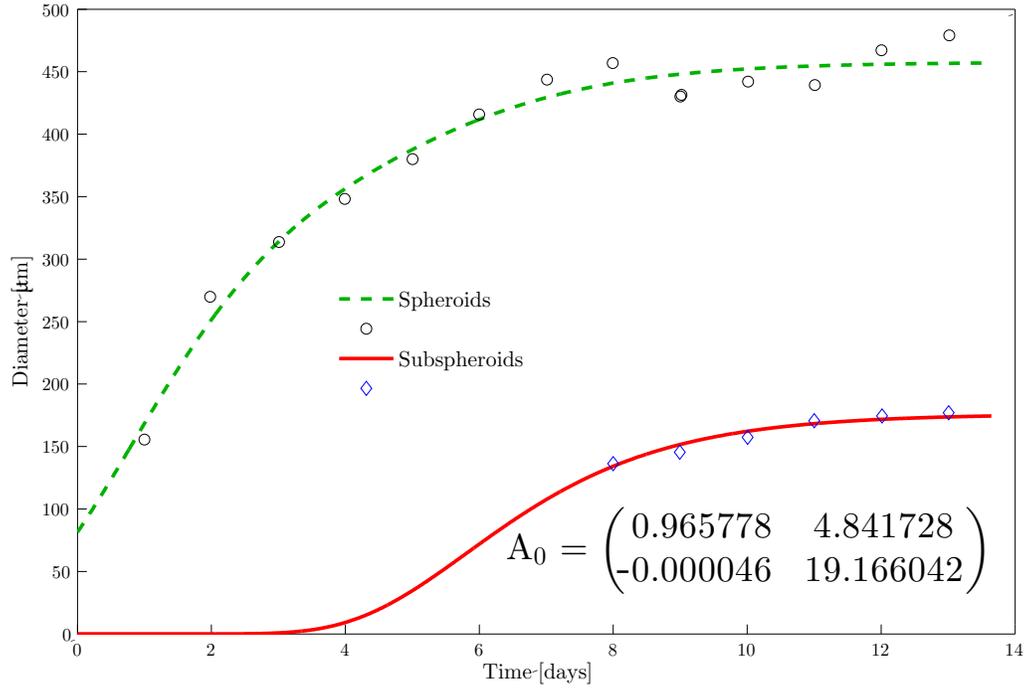}\label{f_sub_fit}
}
\caption{ (a) Images from representative 4T1 tumor breast spheroids on days 4, 6, and 9 of cell culture (data from G\"unther \emph{et al.} (2007), with permission). Subspheroids and isolated detached cells are observed in panel 3. (b) VU1 fit to the spheroid ($y_1$) and subspheroid ($y_2$) populations.\ddag} \label{f_sub}
\end{figure}

\section{Discussion} \label{s_discussion}

Cancer growth is often the result of the competition between two or more cell populations. The Vector Universalities (VUN) formalism introduced in \cite{Barberis2011} and \cite{Barberis2012} can be used to extract information of time series describing the correlated evolution of these populations. We have used this tool to analyze four very different examples of tumor growth, showing that they can be included in three  universality classes and extracting information about various aspects of cancer growth from the observed correlations.

The VU0 class may fit concurrent unrestricted growth. To study an instance of joint growth in this class we measured the simultaneous evolution of two HeLa cell populations (one motile and one reproductive). In this case, it was easy to decide a priori that VU0 was a suitable tool, because we already knew that the colonies were growing in an unbounded and well-nourished environment, a condition that suggests the implementation of the exponential class. A possible difficulty is that when data is scarce multiple good fits may be obtained. In these cases additional information about the system must be taken into account to choose the optimal parameter set. The VU0 tell us why the populations deviate from a purely exponential behavior; they also suggest that a signaling mechanism initiated by the proliferating cells may be responsible for the phenotypic switching of motile cells. Further experimental work is necessary to directly verify this hypothesis.

The VU1 examples in subsections \ref{ss_vitro} and \ref{s_subspherois} are the first to be clearly identified as such in the phenomenological universalities literature.
Here it is important to note that the von Bertalanffy class ($N = 2$), in both the scalar and vector formalisms, adds new degrees of freedom to the fitting problem. If we accept the often used criterion of calculating the $R^2$ value to classify the growth phenomena inside the universality class hierarchy, it is easy to see that in most instances it is possible to find that the VU2 class provides a ``better fit''. In fact, this is the situation in most of the previous PUN work (see \cite{Delsanto2008a} for an example). Because of the VU2 additional degrees of freedom we prefer not to use $R^2$ as a final arbiter when there is no substantial difference between its value for the VU1 and VU2 fits of the same dataset. Instead, in a previous work , we use the Occam's razor criterion as the main argument to classify the phenomena as belonging to the VU1  class \citep{Barberis2012}. Moreover, our knowledge about the biological problem can be used to select the VU2  classification; in example \ref{ss_vivo}, we have 
very good fittings in both universality classes. To assign such a 
system to one or another class is not a simple task and here we do not attempt to do it in general because there is no reliable criterion to define the best class. Instead, our goal is to gather new information for the multipopulation systems from the values of the $A_0$ matrix. For instance, the $A_0 $ matrices recovered in the example in \ref{ss_vivo} give coherent information in both classes, so the classification problem is not really relevant here.

It is interesting that we fail to find any adequate VU2 fitting parameter set for the examples in subsections \ref{ss_vitro} and \ref{s_subspherois}. In the case of subsection \ref{s_subspherois}, we suspect that the absence of an appreciable subspheroid population at early times could play the role of quiescent cells driving the $N=1$ behavior: isolated active cells leave the ``mother'' spheroids in the same way as dying cells leave the live cell population in the MTS problem of subsection 3.2. As it was explained in section \ref{ss_vitro}, the theoretical development of Gompertzian growth is consistent with this description.

As it was pointed out by its creator P.P. Delsanto, the PUN formalism is a novel tool for interdisciplinary and experimental research \cite{Gliozzi2011}. The VUN extension can be used to find very accurate fitting functions for experimental data, as shown in sections \ref{s_pasquale} and \ref{s_subspherois}. It may also be used to test the capability of a theoretical model to reproduce experimental data and define their applicability range, as we did in section \ref{s_MTS}. For cancer researchers, the main advantage of this approach lies in its ability to describe tumor growth using few parameters (whose interpretation is reasonably simple), without making any assumptions on the tissue physiology.

In order to test the reliability and accuracy of the method, the important, and as yet not well understood, problem of the role of the necrotic core in MTS growth was considered. In particular, we analyzed the classical MTS growth data of \cite{Freyer1986}. The result in Fig. \ref{f_MTS_freyer_long}, with a value of $R^2 = 0.9906$ and the prediction of a reasonable saturation mass, verifies the validity of the approach. The similarities and differences between \emph{in vivo} and \emph{in vitro} tumors are highlighted by the similarities and differences between their respective $A_0$ and $\alpha_1$ parameters, as shown in Table \ref{t_MTS}. We remark that the VUN allowed us to obtain excellent fits to the reconstructed data for the evolution of the volumes occupied by the individual (necrotic and live) cell populations. These reconstructed volumes were obtained from a \emph{single} dataset describing the evolution of the total volume of the MTS.

Polyphenols such as baicalein and resveratrol have been shown to increase cell adhesiveness in experimental tumor models \cite{Gunther2007}. Therefore, they can be used to decrease cell detachment and limit the availability of subspheroid seeds, while spheroid coalescence is favored. We predict that, analyzed under the VU1 formalism, the application of these polyphenols should lead to a strong decrease in both $a_{22}$ (fewer seeds due to increased adhesiveness) and $a_{12}$ (the inner cells in the primaries are not exposed to additional 	nutrients due to shedding of the outer shell). VUN fittings to the growth data for the spheroid-subspheroid system could then be used to evaluate the ability of a given polyphenol to decrease the metastatic potential of the cell population under consideration. This ability would be quantified by the matrix elements $a_{12}$ and $a_{22}$.

Summarizing, the implementation of the VUN fitting technique for the analyzed  data sets allow us to conclude that:

\begin{enumerate}

\item  VU0 allows us to better understand and quantify the growth mechanisms of a HeLa cell colony. Phenotype switching is described by the VU0 parameters and resembles the conditions defining the ``go or grow'' hypothesis.

 \item  VU1 confirms that a reconstruction of the (unobserved) MTS necrotic core mass from the total mass data sets is self-consistent and predicts the final spheroid volume starting from a truncated data set. 

\item  VU1 indicates there is a strong similarity between \emph{in vitro} and \emph{ in vivo} cancer growth. It helps us identify quantifiable shared intrinsic features and describe how the different environments affect growth (slowing it down in the in vivo samples).

\item  VU1 validates the consistency of the hypothesis that necrotic cells generate signals that (a) restricts growth and (b) kills live cells.  

\item  VUN leads to a better understanding of the interaction between spheroids and subspheroids and could help to characterize the effect of polyphenols on cancer cell shedding and dispersal.

\end{enumerate}

 As a final remark we would like to emphasize that the formalism used in this contribution may also be applied to the analysis of other datasets related to ecological management and population dynamics. Examples on these topics were provided in \cite{Barberis2012}. Further developments to include more than two populations, seasonality, and external forcing are in progress.\linebreak

{\bf Acknowledgements}\linebreak

 This work was supported by SECyT-UNC (Project 30720110100436) and CONICET (PIP 00772 and PIP 02231), Argentina.

 \appendix

\section{Theoretical examples}

\begin{figure} 
\subfloat[][]{
 \includegraphics[width=0.5\textwidth]{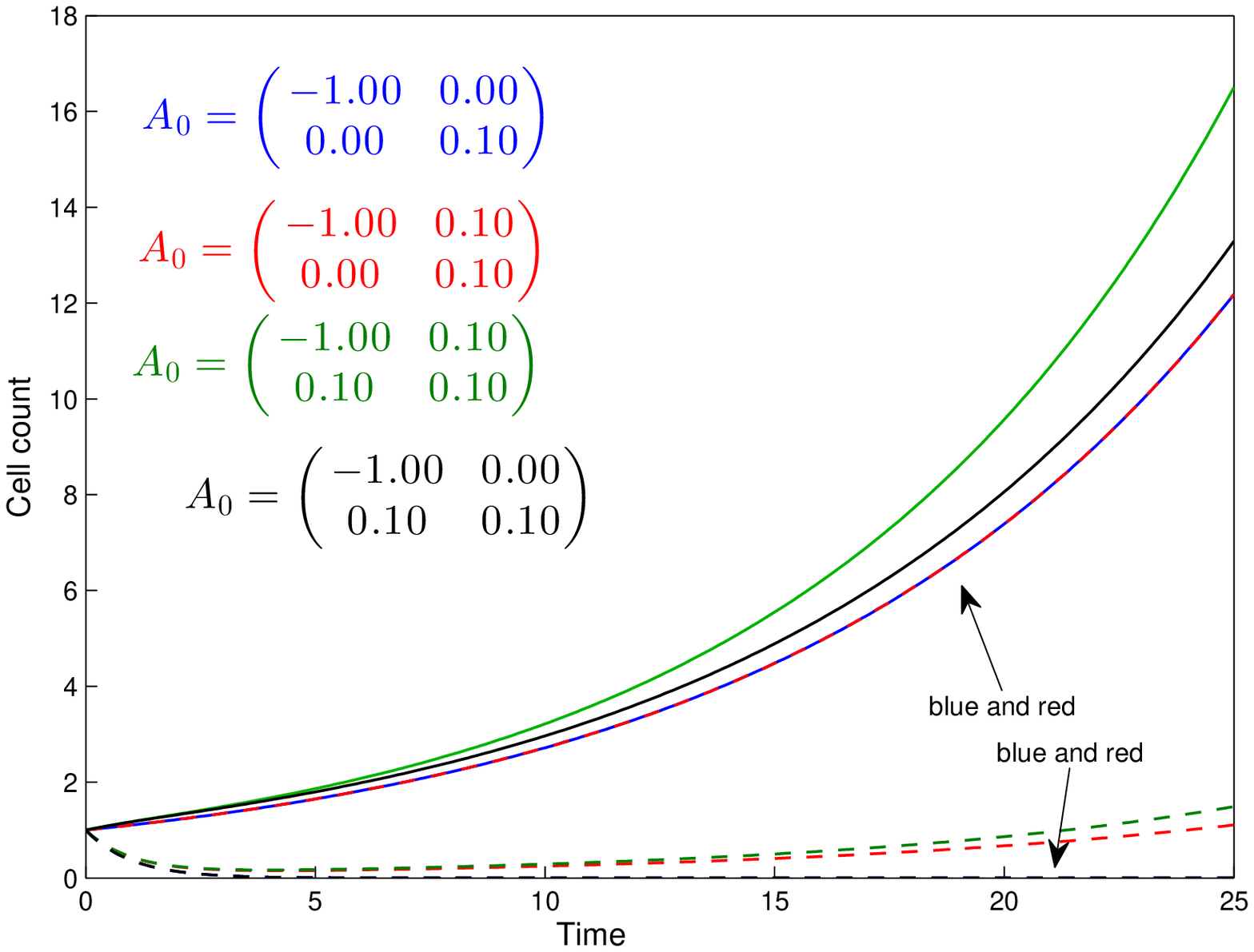}
\label{f_interplay_ea}}
\subfloat[][]{
\includegraphics[width=0.5\textwidth]{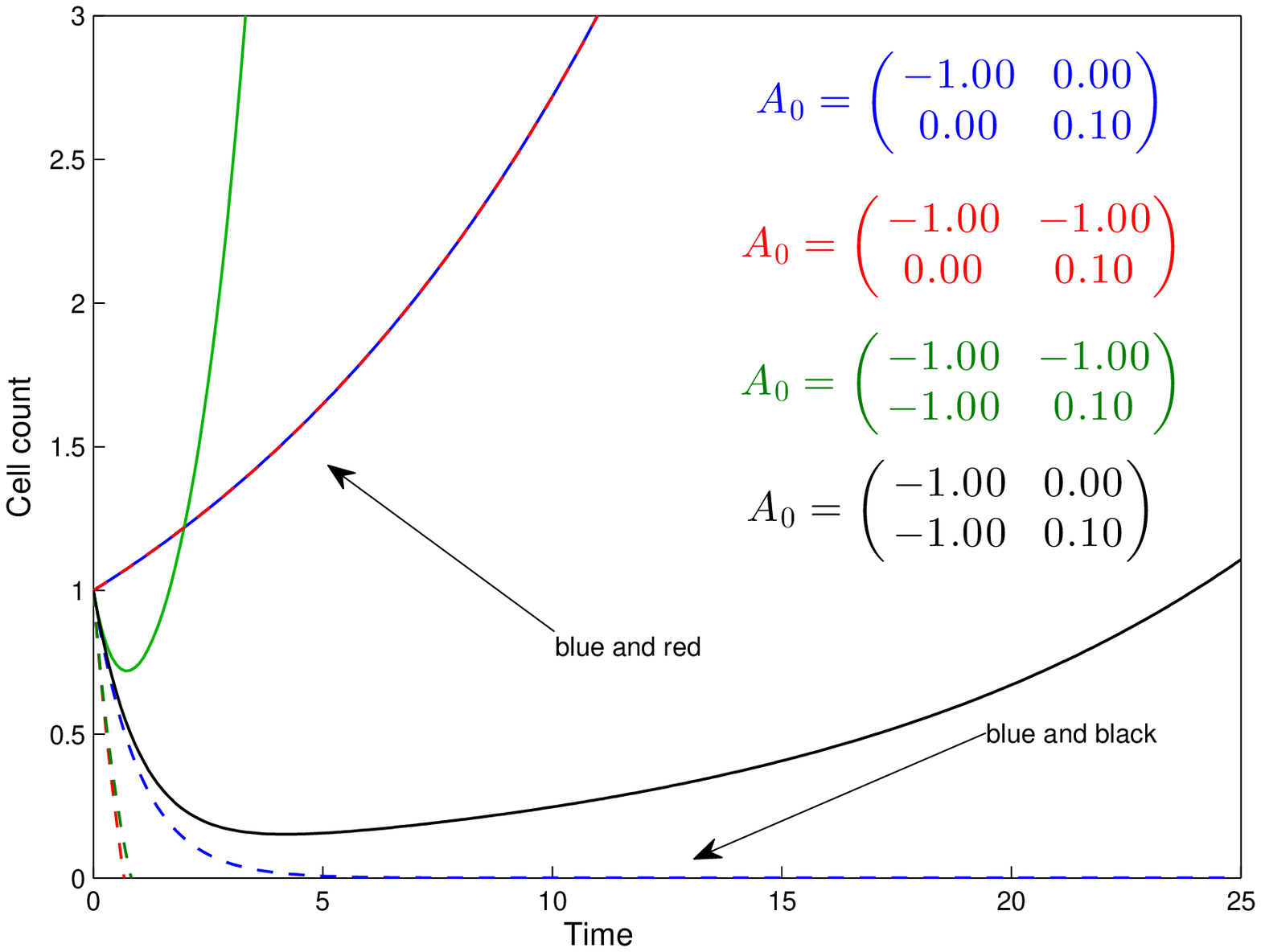}
\label{f_interplay_eb}}\\
\centering
\subfloat[][]{
\includegraphics[width=0.5\textwidth]{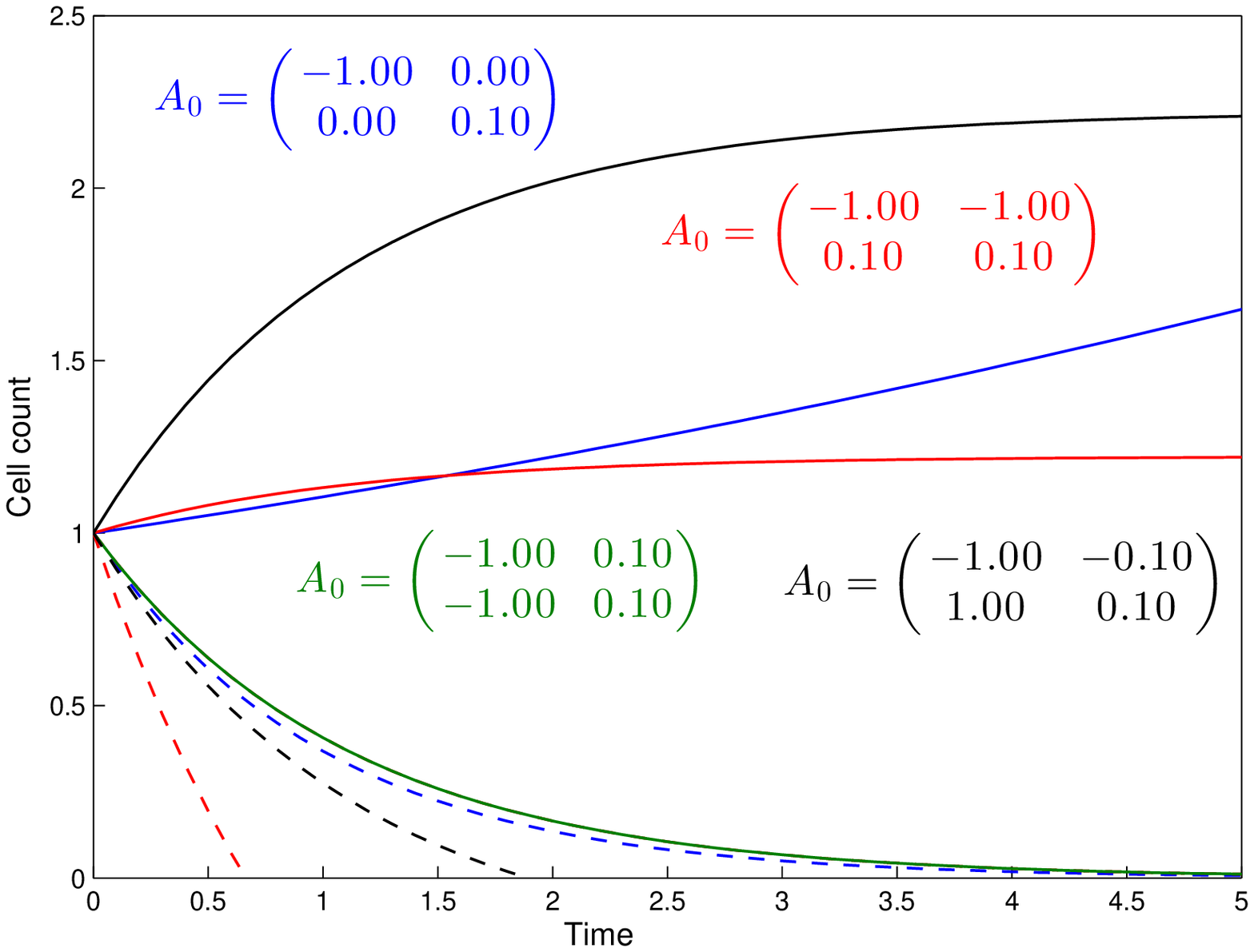}
\label{f_interplay_ec}}\caption{Interplay between two agents, represented by solid and dashed lines, respectively, in the VU0 class  for various  choices of the $A_0$ matrix.  Blue matrices and lines represent a reference situation without direct interactions. Other matrix choices may be used to describe such processes as (a) cooperation, (b) competition, and (c) parasitism. Each agent pair is described by one solid and one dashed line. More details in the text.  }\label{f_interplay_e}
\end{figure}

\begin{figure} 
\centering
\subfloat[][]{
 \includegraphics[width=0.7\textwidth]{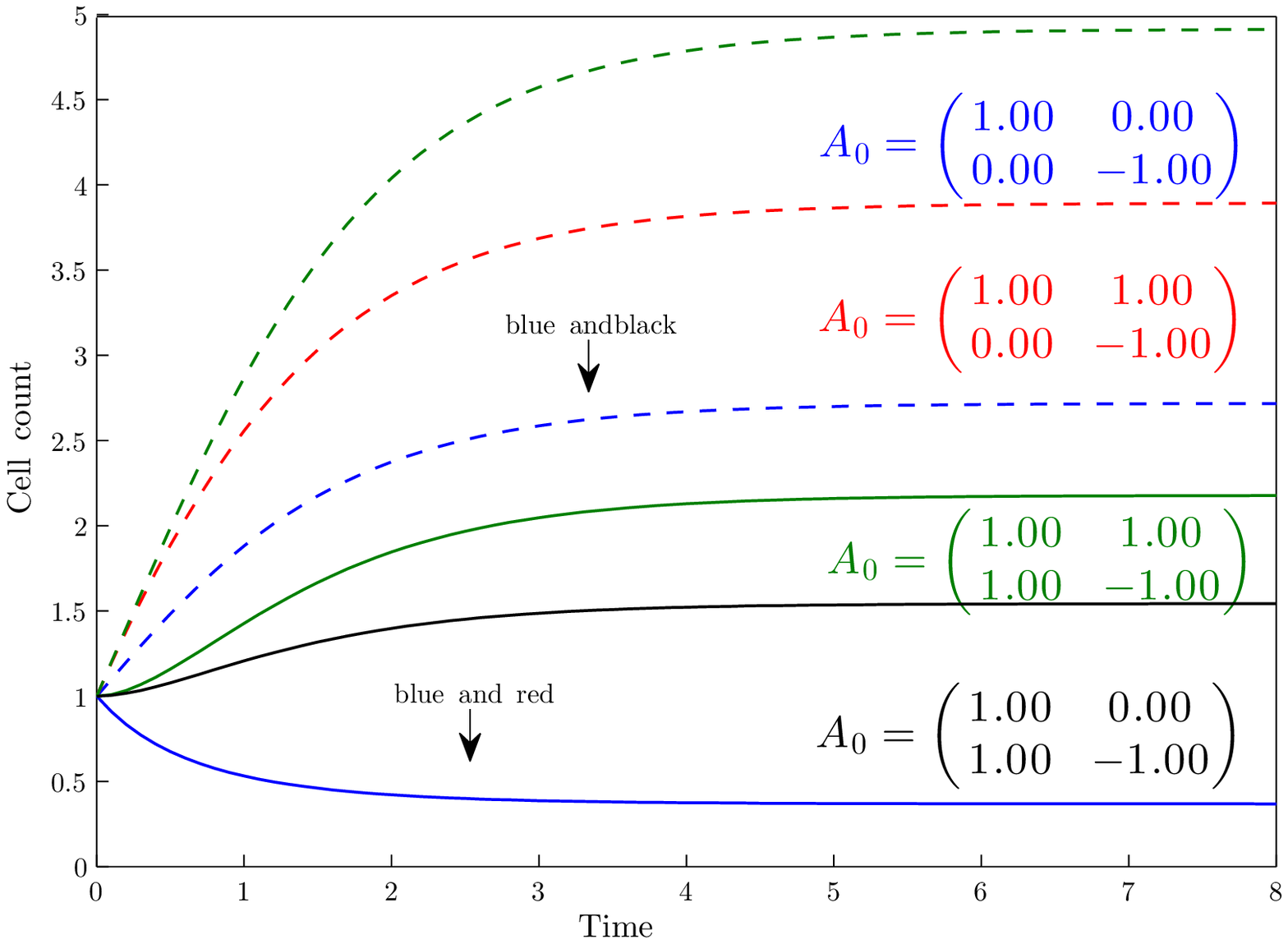}
}\\
\subfloat[][]{
\includegraphics[width=0.7\textwidth]{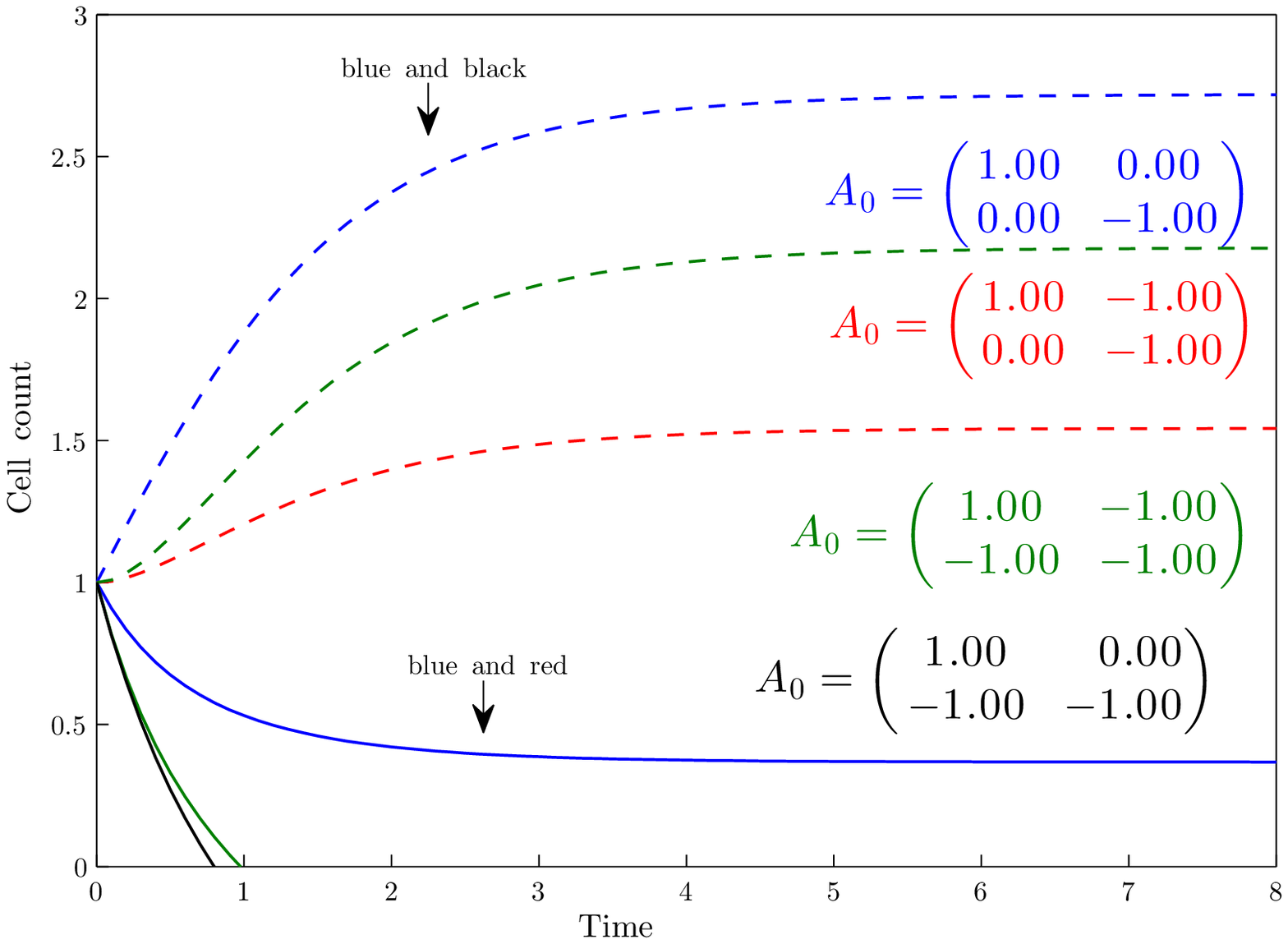}
}
\caption{Interplay between two agents in the VU1 class for various choices of the $A_0$ matrix. (a) Cooperation between both species (green) and one species helping another (red and black). (b) Competition (green) and parasitism (red and black).
Each agent pair is described by one solid and one dashed line.  }\label{f_interplay_g}
\end{figure}

The behavior of the functions belonging to the VU0 class is regulated by a linear combination of the exponential functions $f_0(\lambda_j;t)=exp(\lambda_j t)$. As a benchmark, we have chosen  a two-agent system and a trait that would evolve according to a diagonal  $A_0$ matrix whose elements have opposite signs. Such a matrix corresponds to agents that only influence each other directly, i.e., without modifying their own environment. We show in Figures \ref{f_interplay_e} and \ref{f_interplay_g} several examples of interactions. Blue lines and blue matrices correspond to the reference  situation. Note that the first individual lives in a very hard environment ($a_{11}$ large and negative), while the second one has a moderately friendly environment ($a_{22}$ small and positive). As a consequence the ``size'' of the trait in the first agent decreases, disappearing after 5 time units, while that of the second agent exhibits continuous growth.

\textit{Exponential growth}: Figure \ref{f_interplay_e} shows several examples that illustrate  how the reference situation is changed by the addition of off-diagonal elements. Figure \ref{f_interplay_ea} describes cooperation when there is an extra gain represented by positive off-diagonal elements. Note how what was a disappearing trait/population represented by the blue dashed line the reference case, picks up growth due to the interaction and, as a consequence,  the trait/population survives. Figure \ref{f_interplay_eb} shows the case mentioned in section \ref{s_pasquale}: the annihilation of one population  (dashed line) or the regrowth of an initially decaying population (continuous green) in the presence of competition.  Note that in the case described by the dashed red line annihilation occurs five times faster. On the other hand, Fig. \ref{f_interplay_ec} shows an example of parasitism (opposite signs of the off-diagonal elements) leading to different survival times. Of course there is a 
manifold of matrix element 
combinations that would allow us to describe many possible situations.

\textit{Gompertzian growth}: As in the previous exponential case, we show in Fig. \ref{f_interplay_g} several examples of the possible behavior of the VU1 system. Cooperation, competition, and parasitism can be addressed by a suitable choice of the off-diagonal elements of $A_0$. The diagrams are self-descriptive.   
\pagebreak

\bibliographystyle{model2-names}
\bibliography{biblio.bib}

{\small
\noindent \dag ~Reprinted by permission from Macmillan Publishers Ltd: \emph{Nature}, Han, M \emph{et al.},`` Enhanced Percolation and Gene Expression in Tumor Hypoxia by PEGylated Polyplex Micelles'' {\bf 17}(8), copyright (2009) 

\noindent \ddag ~Reprinted from \emph{Cancer Lett.}, {\bf 250}(1), G\"unther, S \emph{et al.}, ``Polyphenols prevent cell shedding from mouse mammary cancer spheroids and inhibit cancer cell invasion in confrontation cultures derived from embryonic stem cells'', 25, Copyright (2007), with permission from Elsevier.}
\end{document}